\documentclass[12pt]{iopart}
\usepackage{graphicx}

\newcommand{\tmr}{\frac{2M}{R}}
\newcommand{\rx}{\frac{r}{R}}

\newcommand{\om}{\omega}

\begin{document}

\title{Gravitational waves from instabilities in relativistic stars}

\author{Nils Andersson}

\address{Department of Mathematics, University of Southampton, Southampton 
SO17 1BJ, UK}

\begin{abstract}
This article provides an overview of stellar instabilities as sources
of gravitational waves. The aim is to put recent
work on secular and dynamical instabilities in compact stars in context, 
and to 
summarize the current thinking about the detectability of gravitational
waves from various scenarios. As a new generation of 
kilometer length interferometric detectors are now coming online 
this is a highly topical theme.
The review is motivated by two key questions for future
gravitational-wave astronomy: Are the gravitational waves from 
various instabilities detectable? If so, what 
can these gravitational-wave signals teach us about neutron star physics?
Even though we may not have clear 
answers to these questions, recent studies of the dynamical 
bar-mode instability and the secular r-mode instability have provided 
new insights into many of the difficult issues involved in 
modelling unstable stars as gravitational-wave sources.
\end{abstract}

\section{Introduction}

Neutron stars may suffer a number of instabilities. These instabilities
come in different flavours, but they have one general feature in common: 
They can be directly associated with unstable modes of oscillation. 
A study of the stability properties of a relativistic star is
closely related to an investigation of the star's various pulsation modes. 
Furthermore, non-axisymmetric stellar oscillations 
will inevitably lead to the production of gravitational radiation.
Should these waves turn out to be detectable,
they would provide a fingerprint that could  be used to 
put constraints on the interior structure of the star \cite{astero}. 
This would 
be analogous to the recent success story of helioseismology, where
the detailed spectrum of solar oscillation modes has been matched to 
theoretical models of the  interior to provide insights into, 
for example, the sound speed at different depths in the Sun. 
In order for ``gravitational-wave 
asteroseismology'' to be a realistic proposition, one must 
find scenarios  which lead to a star pulsating wildly.
The most obvious situation where this may be the case is 
when a newly born neutron star settles down after 
the supernova collapse. Other promising possibilities are
associated with instabilities. As an unstable pulsation mode grows
 it may reach a sufficiently
large amplitude that the emerging gravitational waves can be detected.

The aim of this article is to provide an 
overview of instabilities that may lead to 
detectable gravitational waves. This is a topical 
theme given that several large-scale interferometers (LIGO, GEO600, VIRGO and TAMA300)
will be up and running in the near future.
My intention is to put the current thinking about various 
stellar instabilities in context, and provide a
foundation for future research in this area by summarizing
what we currently know and, perhaps more importantly, what the 
key issues that need further attention are.
This review is, however, by no means 
exhaustive. It does not provide a ``complete'' 
set of references to the literature or a 
detailed discussion of every possible instability that may occur in a 
relativistic star. I have simply focussed on the issues that seem (to me)
to be the most interesting/important. The reader is  encouraged to 
use the provided bibliography as an entry point to the vast 
literature on instabilities and other 
gravitational-wave sources \cite{bfs1,bfs2,kip,cutler,hughes}, 
explore it for him/herself, and perhaps even draw 
conclusions that differ from mine.

Neutron stars are tremendously complicated objects. In essence, their 
modelling requires a detailed understanding of the very extremes of physics, including
supranuclear physics, general relativity, 
superfluidity/superconductivity, strong magnetic fields, exotic particle 
physics etcetera \cite{glend}. To investigate the pulsation properties
of any ``realistic'' neutron star model is therefore a serious challenge.
While our understanding of the modes of a simple self-gravitating 
``ball of fluid'' may be good, a significant effort is still 
required if we want to model the dynamics of astrophysical compact stars.       

The instabilities that I will discuss fall into two main
categories: They are either dynamical or secular. In the first case
the unstable modes grow on a timescale similar to that of the oscillations,
while in the second the growth takes place on a much longer timescale
(eg. that associated with viscosity).
A dynamical instability 
is likely to have dramatic effects on an equilibrium configuration, 
while a secular instability acts in a more subtle way. 
As we will see, this means that dynamical instabilites can 
be studied directly using fully nonlinear hydrodynamical simulations.
The effects of a secular instability are difficult to explore in this 
way since the evolutionary timescale tends to be much longer than the
dynamical timescale of the system. Hence,  secular instability evolutions
have so far mainly been discussed at a phenomenological level.

Most of the discussion in this article concerns rotating stars. 
This is natural
since  instabilities in spinning stars are thought
to be particularly 
promising gravitational-wave sources. A question of key 
astrophysical importance concerns whether  instabilities may limit 
the spin of a compact star to a rate significantly  lower than 
the mass-shedding limit. This limit is reached when the equator
rotates at the Kepler frequency of a particle in circular orbit 
around the star. It is well approximated by
a rotation frequency
\begin{equation}
\Omega_K \approx { 2 \over 3} \sqrt{ \pi G \rho_0} 
\label{kepler}\end{equation}
where $\rho_0$ is the average density of the corresponding 
non-rotating star. 
This is known to be a good approximation for rigidly rotating Newtonian
polytropes \cite{il1}, 
and it remains reasonably accurate also for 
relativistic models \cite{nslr}. However,  we should keep in mind that the 
mass-shedding limit 
may be significantly different in the case of differentially rotating stars.
In fact, the rotation frequency $\Omega$ is not a particularly useful parameter
for differentially rotating configurations (see \cite{tassoul}). Instead, 
 the ratio $\beta$ between the kinetic energy $T$ and the 
gravitational potential energy $W$ is often used in discussions of rotational instabilities.
The values of $\beta$ are restricted (by the virial theorem)
to the range $0-0.5$.
For a uniformly rotating constant density star one can estimate that 
\begin{equation}
\beta = { T \over |W|} \approx
{ 1 \over 9} \left( {\Omega \over \Omega_K} \right)^2
\label{betadef}\end{equation}
In other words, the mass shedding limit would correspond to 
$\beta_K \approx 0.11$. This simple approximation agrees
quite well with detailed calculations for 
realistic supranuclear equations of state\footnote{For a detailed review of various 
methods used to construct stationary rotating stellar models in General 
Relativity, see \cite{nslr}.}, which typically
indicate a maximum value of $\beta$ in the range 0.09-0.13.
Depending on the rotation law, differentially rotating models
may allow considerably larger values of $\beta$ (maybe as large as 
$\beta \approx 0.3$). Large maximum values of $\beta$ can also be reached 
for strange stars. Such self-bound objects may have
$\beta$ significantly larger than 0.2 \cite{gour}.

\section{Gravitational-wave estimates}
\label{Sec:gw}

Among astrophysicists General Relativity is sometimes viewed as a 
``correction'' (expected to alter results quantitatively 
at the 15-20\% level). This attitude makes some sense since
many astrophysical results remain essentially unaltered by a 
fully relativistic description. On the other hand, there are  
 situations where General Relativity is a
``leading order effect'' which must  
be incorporated. The dynamics of
compact stars provides an excellent example of this. 
It is well known that it is meaningless (or at least 
highly dubious) to use a realistic supranuclear equation of 
state in a Newtonian calculation of neutron star structure. 
The mass and radius of the resultant star may differ greatly
from the results obtained from the Tolman-Oppenheimer-Volkoff
equations (for a given central density), and in order to avoid
confusion one should always use relativistic models for
``realistic'' neutron stars. In addition, gravitational waves 
are a purely relativistic phenomenon. So if we are interested 
in neutron stars as gravitational-wave sources it stands to reason that 
we should aim to build fully relativistic models. 
However, this is a far from simple task. 
Since much of the  physics required for a realistic model of neutron star 
dynamics is poorly understood  we are in practice
often forced to work with Newtonian models, estimating gravitational waves
via post-Newtonian formulas.

If we focus our attention on a single pulsation mode of a 
rotating Newtonian star then, assuming that the density and velocity 
perturbations,
$\delta \rho$ and  $ \delta \vec{v}$, 
depend on time as $\exp(i\omega_r t)$, 
the associated gravitational-wave luminosity
(measured in the rotating frame) can be estimated using \cite{thorne}
\begin{equation}
 {dE\over dt} = - \omega_r \sum_{l=2}^\infty N_l
\omega_i^{2l+1} \left( | \delta D_{lm} |^2 + | \delta J_{lm} |^2 \right) \ ,
\label{gwlum}\end{equation}
where $\omega_i$ ($\omega_r$) is the mode-frequency in the inertial (rotating) 
frame, and
\begin{equation}
N_l = {4\pi G \over c^{2l+1} } { (l+1)(l+2) \over l(l-1)[(2l+1)!!]^2 } \ .
\end{equation}
The first term in the bracket of (\ref{gwlum}) represents radiation due to
the mass multipoles. These are determined by
\begin{equation}
\delta D_{lm} = \int \delta \rho r^l Y_{lm}^* dV \ .
\label{mass}\end{equation}
(where the asterisk represents complex conjugation). 
The second term in the bracket of (\ref{gwlum})  corresponds to
the current multipoles, which follow from 
\begin{equation}
\delta J_{lm} = {2 \over c} \sqrt{ {l\over l+1}} 
\int r^l (\rho \delta \vec{v} +
\delta \rho \vec{\Omega}) \cdot \vec{Y}_{lm}^{B*} dV \ ,
\label{curr}\end{equation} 
where 
\begin{equation}
\vec{Y}_{lm}^{B} = { 1 \over \sqrt{l(l+1)}} \hat{r} \times \nabla Y_l^m 
\end{equation}
are the magnetic multipoles \cite{thorne}.

From the above formulas we can draw some general conclusions. 
First of all it is clear that any fluid motion that leads to 
significant density variations will radiate gravitationally
predominantly through the mass multipoles. This follows 
from the fact that $|\delta J |^2\sim |\delta D |^2/c^2$ which means that
the current multipole radiation
is generally ``one order higher'' in the post-Newtonian approximation\footnote{By counting
the inverse powers of $c$ in each of the terms in (\ref{gwlum}) we see that quadrupole
perturbations ($l=2$) will lead to $\dot{E}\sim 1/c^5$ for the mass multipoles, while
$\dot{E}\sim 1/c^7$ for the current multipole radiation. 
Using the standard way of counting orders, this means that the
mass multipole radiation arises at 2.5 post-Newtonian order while the current 
multipole radiation is a 3.5 post-Newtonian effect.}.
However, there are situations where this standard 
consensus no longer holds and the current multipoles 
provide the main radiation mechanism. Most notably, this is
the case for the unstable r-modes which are
characterized by 
a large $\delta \vec{v}$ and a small $\delta \rho$. 

In the relativistic case, the calculation of  gravitational-wave 
emission from stellar pulsation is conceptually more
straightforward. Because any non-axisymmetric fluid motion 
will lead to radiation, the various modes are  distinguished by 
imposing outgoing-wave boundary conditions
at infinity. This means that they are no longer normal modes of the system.
Gravitational-wave dissipation leads to  the mode-frequencies 
becoming complex (with the imaginary part representing
the damping/growth due to the emitted radiation). 
As is well-known from studies of perturbed black holes
the numerical determination of such ``quasinormal modes''  
is not trivial. Several reliable methods for 
 handling this difficulty for spherical 
stars have been developed \cite{qnm}, but a solution to the problem 
for rapidly rotating stars is still outstanding.

Once we have some idea of the character of a given gravitational-wave
source it is relevant to try to estimate the detectability of the 
signal. The obtained estimates are often rough, but they  
provide useful guidelines for further work, both on the theoretical
modelling side and the experimental side. If initial 
estimates make it seem plausible that the source can be detected
then a detailed data analysis strategy needs to be developed. 
In the ideal case this strategy would be based on
matched filtering, where  an accurate theoretical template is 
used to search for a
real signal in the noisy data stream. 
However, in many cases our understanding of the relevant 
physics is not at the level where we can expect to provide 
reliable theoretical models. This is certainly the case for 
neutron stars, where crucial pieces of physics like the 
equation of state are only known to within a factor of two or so.
The detectability of a signal can nevertheless be improved considerably
given only overall characteristics, like the  duration and  
the frequency range. Any information we can extract from 
the theoretical models, even if they are rudimentary, could be valuable!

In order to estimate the strength of a signal, we can use the well-known flux formula
which relates the luminosity to the gravitational-wave strain $h$;
\begin{equation}
{ c^3 \over 16 \pi G} |\dot{h}|^2 = { 1 \over 4 \pi D^2} { dE \over dt}
\end{equation}
where $D$ is the distance to the source. This relation is exact for 
the weak waves that bathe the Earth. 
To proceed we characterize the event by a timescale
$\tau$ and assume that the signal is essentially monochromatic. 
Then we can use $\dot{h} \approx 2 \pi f h$, and readily 
deduce that
\begin{equation}
h \approx 5\times 10^{-22} \left( { E \over 10^{-3} M_\odot c^2} \right)^{ 1/2}
\left({ \tau \over 1 \mbox{ ms} }\right)^{-1/2} f_{Hz}^{-1} D_{15}^{-1}
\label{hraw}\end{equation}
Here we have taken the  distance to be that of a source
in the Virgo cluster. This is necessary to ensure a reasonable event rate 
for most astrophysical scenarios. At that distance one would expect to
see many supernovae per year, which means that one can 
hope to see a few neutron stars being born during one year of
observation.

\begin{table}
\caption{Canonical parameter values used in various formulas in the article.}
\begin{tabular}{lll}
$M_{1.4}$ & $M/1.4M_\odot$ & mass \\
$R_{10}$ &  $R/10\mbox{ km}$ & radius \\
$P_{-3}$ &  $P/1~\mbox{ms}$ & rotation period \\ 
$T_9$ & $T/10^9~\mbox{K}$ & core temperature \\
$\rho_{15}$ & $\rho / 10^{15} {\rm g/cm}^3$ & density\\
$f_{Hz}$ & $f/{\rm 1}$~Hz & gravitational-wave frequency \\
$D_{15}$ & $D/15$~Mpc & distance to source
\end{tabular}
\label{param}\end{table}

We can proceed further and estimate an ``effective amplitude'' that 
reflects the fact that detailed knowledge of the signal can be used to 
dig deeper into the noise. A typical example is based on the use of matched
filtering, for which the effective amplitude 
improves as the square root of the number of observed cycles $n$. 
Using $n \approx f \tau $ we arrive at 
\begin{equation}
h_{\rm c} 
\approx 5\times 10^{-22} \left( { E \over 10^{-3} M_\odot c^2} \right)^{ 1/2}
\left({ f \over 1 \mbox{ kHz} }\right)^{-1/2} D_{15}^{-1}
\label{heff1}\end{equation}
From this relation we see that
the ``detector sensitivity'' essentially depends only on the 
radiated energy and the characteristic frequency. 
Hence, an estimate of the total energy radiated and the
frequency of the signal may be sufficient to assess
the relevance of the event as a gravitational-wave source.

The above formulas can be applied to many interesting astrophysical
scenarios, but there are obvious situations where they fail. 
One case, that will be discussed later, is when gravitational
radiation reaction leads to a change in the character of the signal 
during the observation. This typically leads to a signal whose 
frequency varies with time. 
Provided that this variation is 
sufficiently slow one can use the method of stationary 
phase to show that
\begin{equation}
h_{\rm c} \approx h \sqrt{ f^2 \left| { dt \over df } \right| } 
\label{heff2}\end{equation}
For example, Owen et al \cite{owen} used
this relation to assess the detectability of 
gravitational waves from unstable r-modes. 

\section{Stellar pulsation} 

In this Section I will describe the nature of those
pulsation modes that are currently thought to be the most relevant from
the gravitational-wave point of view. 
The main focus will be on the acoustic f-modes and the inertial r-modes, 
but I will also briefly mention the gravitational-wave 
w-modes. For more details, the reader is refered to standard
textbooks \cite{tassoul,unno} and various 
review articles  \cite{akreview,lindrev,lockrev}. 
 
\subsection{The equation of state}

In order to close the system of equations that describe linear
oscillations of a star it is necessary to provide an equation of state 
for matter. This problem is extremely difficult --- at least if we expect 
to find the ``correct'' answer. 
The proposed equations of state for supranuclear matter come in 
many varieties and lead to equilibrium neutron stars whose bulk properties 
vary by up to factors of two. The available predictions also 
differ considerably on issues concerning the nature of matter at extreme
densities. Many important questions remain to be resolved. Will a neutron star
have an exotic core containing deconfined quarks and/or hyperons?
Are there strange matter stars? 
At what density and temperature do various constituents become 
superfluid/superconducting? As we will discuss later, the answers
to these questions are crucial for an understanding 
of the stability properties of relativistic stars. At the time 
of  writing,  the permissible parameter space is enormous.

Detailed nuclear physics calculations typically provide a
tabulated equation of state relating the pressure to the density
for matter in beta equilibrium. This provides sufficient information for us to 
construct eg. rotating equilibria, and thus model
unperturbed neutron stars. In these models
matter is usually modelled as a perfect fluid. 
The temperature is typically (unless one considers newly born
neutron stars) 
taken to be zero, because even though neutron stars may seem 
hot on a ``normal physics scale'', 
with interior temperatures in the range $10^6-10^8$~K,
the thermal energy is considerably smaller than the Fermi energy
of the  fluid, $T_{\rm Fermi} \sim 10^{12}$~K. 
As a consequence a one-parameter equation of state
$p=f(\rho)$ is often sufficient to
describe the matter. In this article  stars 
constructed from such equations of state are refered
to as ``barotropic''.
A simple  yet reasonable class of equations
of state are the polytropes 
\begin{equation}
p = \kappa \rho^\Gamma \ , \quad \mbox{ with} \quad \Gamma = 1 + { 1\over n}
\label{poly}\end{equation}
where $n$ is the polytropic index. 
A comparison with proposed realistic equations of 
state suggests that $\Gamma \sim 2$, and hence 
model calculations are often carried out for 
$n=1$ polytropes.  

More detailed models account for the dependence on additional
parameters, the most important of which may be the 
relative number density of protons ($x_p = \rho_p/\rho$)
and exotic particles like hyperons and/or deconfined quarks \cite{glend}. 
Stars described by such multi-parameter equations of state, 
$p=f(\rho,x_p,...)$, are ``non-barotropic'' and
have internal stratification. 
This stratification may affect the pulsation properties of the
star significantly. For example, internal stratification associated 
with chemical composition gradients leads to the presence
of distinct gravity  g-modes \cite{rg}. 
 
Since we want to study oscillating stars we face a further challenge. 
The mode motion will inevitably force the fluid out of equilibrium, 
which means that the available tabulated equations of state 
only provide  partial information. To address the 
pulsation problem we need to know the ``exact'' equation of state
in order to relate the various (Eulerian) perturbations, eg.
\begin{equation}
\delta p = \left. { \partial p \over \partial \rho} \right|_{x_p,...} \delta \rho 
+  \left. { \partial p \over \partial x_p } \right|_{\rho,...} \delta x_p 
+  ...
\end{equation}
This requires the knowledge of physics that is poorly understood. 
Hence it is customary to approach the problem in the following way: 
We simply allow the perturbations to be governed by an equation of state 
that differs from that which describes the background configuration. 
To do this we relate the Lagrangian pressure and density variations by
\begin{equation}
\Delta p = { \Gamma_1 p \over \rho} \Delta \rho 
\end{equation}
where $\Gamma_1$ need not be equal to $\Gamma$. 
From this definition it immediately follows that the Eulerian variations
are related by
\begin{equation}
{\delta p \over p} = \Gamma_1 {\delta \rho \over \rho} + \Gamma_1
\vec{\xi} \cdot \left[ \nabla \log \rho -
{1 \over \Gamma_1 }\nabla  \log p  \right] =  \Gamma_1 {\delta \rho \over \rho} + 
{\vec{\xi} \cdot  \nabla\rho \over \rho} (\Gamma_1 -\Gamma) 
\label{eospert}\end{equation}
where $\vec{\xi}$ is the Lagrangian displacement vector , i.e. 
$\delta \vec{v} = \partial_t \vec{\xi}$, 
and the last equality only holds for polytropes.
Here it is customary to introduce the so-called 
Schwarzschild discriminant ${\cal A}_s$. For a spherical 
stellar model where  $p$ and $\rho$ 
depend only on the radial coordinate $r$  we have
\begin{equation}
{\cal A}_s = {1 \over \rho} {d\rho \over dr} -{1 \over \Gamma_1 p} 
{dp \over dr}  = {\Gamma_1 - \Gamma \over \Gamma_1} {1 \over \rho} {d\rho \over dr} 
\end{equation}
The key point of this approach
is that one can easily account for the buoyancy due to internal 
stratification (whatever the physical reasons for its presence
may be) 
by considering models with a non-zero Schwarzschild discriminant.
This way we can parameterize our ignorance concerning the detailed
physics involved in real stellar pulsations.

\subsection{f- and r-modes}

A realistic neutron star model has a large number of 
families of pulsation modes with more or less distinct character. 
In principle, one expects each physical ``restoring force'' that
acts on a fluid element to lead to the presence of a new
family of pulsation modes. 
An exhaustive description of the all these various modes 
goes far beyond the scope of the present article. 
Here we are mainly interested in modes that may become unstable and
lead to a significant gravitational-wave signal. The modes that are 
currently thought to be the most important in this respect are
the acoustic f-modes and the Coriolis restored r-modes.

In addition to being associated with different restoring forces, 
stellar pulsation modes can be classified by the way that the 
perturbations transform under parity. 
This is a useful classification 
since the overall parity of a perturbation  is preserved 
when rotation is 
present. This means that one can follow the changes in a particular
parity mode as $\Omega$ is varied, and meaningfully label 
the mode by its nature in the non-rotating limit (eg. the 
spherical harmonic indices $l$ and $m$ used to describe the 
angular dependence and the number of radial nodes of the
mode-eigenfunction). 

A general velocity field can be represented by 
\begin{equation}
r \delta \vec{v} = \sum_{lm} \left\{ W_l^m \vec{r} Y_l^m + V_l^m r \nabla Y_l^m -
i U_l^m \vec{r}\times \nabla Y_l^m \right\}
\label{vdec}\end{equation}
where $Y_l^m(\theta,\varphi)$ are the standard spherical harmonics. 
The first two terms in the sum 
transform as $(-1)^l$ under parity,
 while the last term changes as
$(-1)^{l+1}$.  These two classes of perturbations  are often (in 
relativity) referred to as polar and axial \footnote{In Newtonian 
stellar pulsation studies the two classes of modes are often called
``spheroidal'' and ``toroidal''.}, respectively.
This nomenclature was introduced by Chandrasekhar
for perturbed black holes \cite{Chandrabook}. 
In a non-rotating star the various multipoles decouple, but in the rotating 
case they are coupled in such a way that the overall parity of the 
mode is preserved. That is, polar terms corresponding to even values of $l$ are
coupled to axial terms for odd $l$, and vice versa. 

\begin{figure}
\centering
\includegraphics[height=8cm,clip]{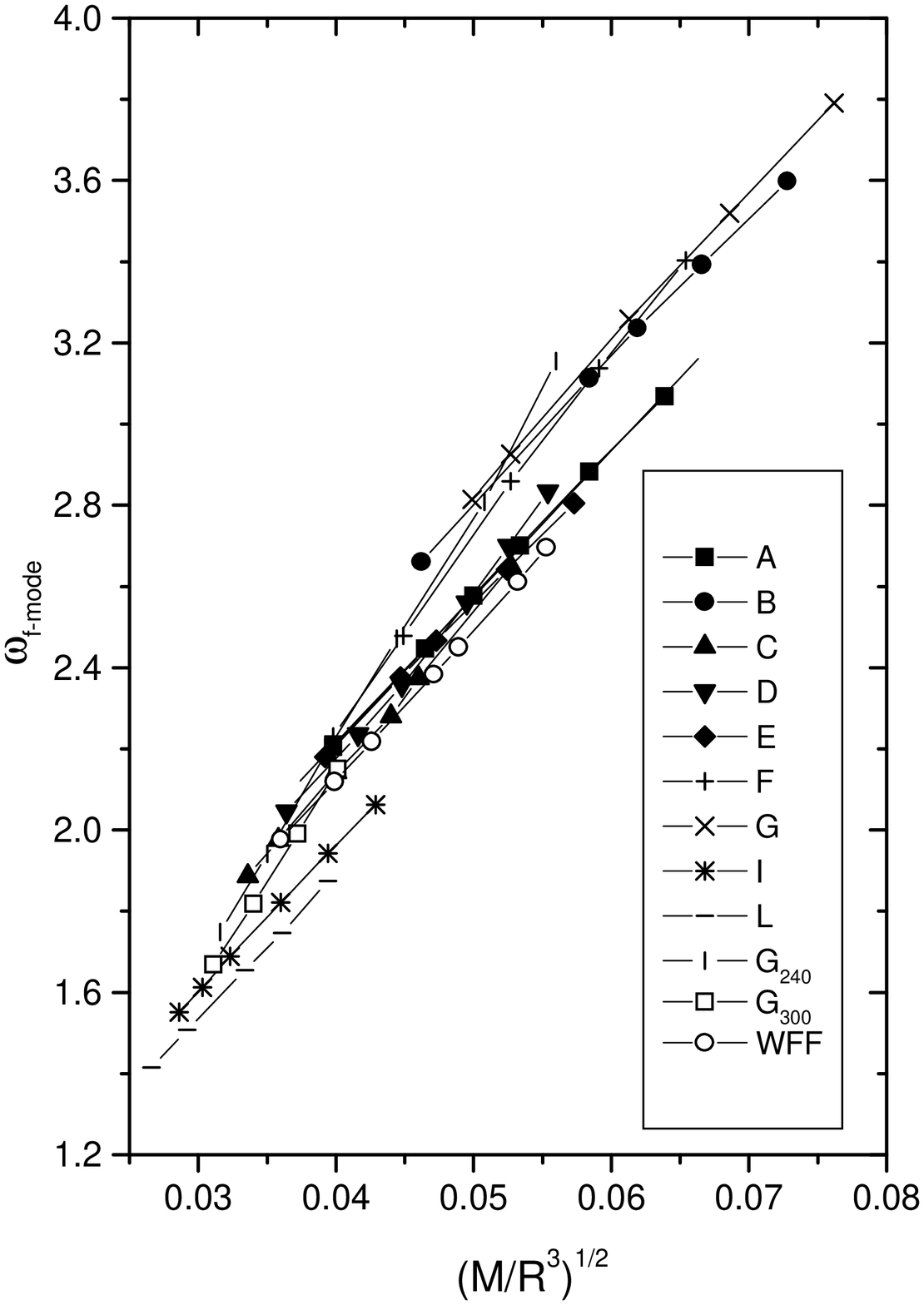}
\includegraphics[height=8cm,clip]{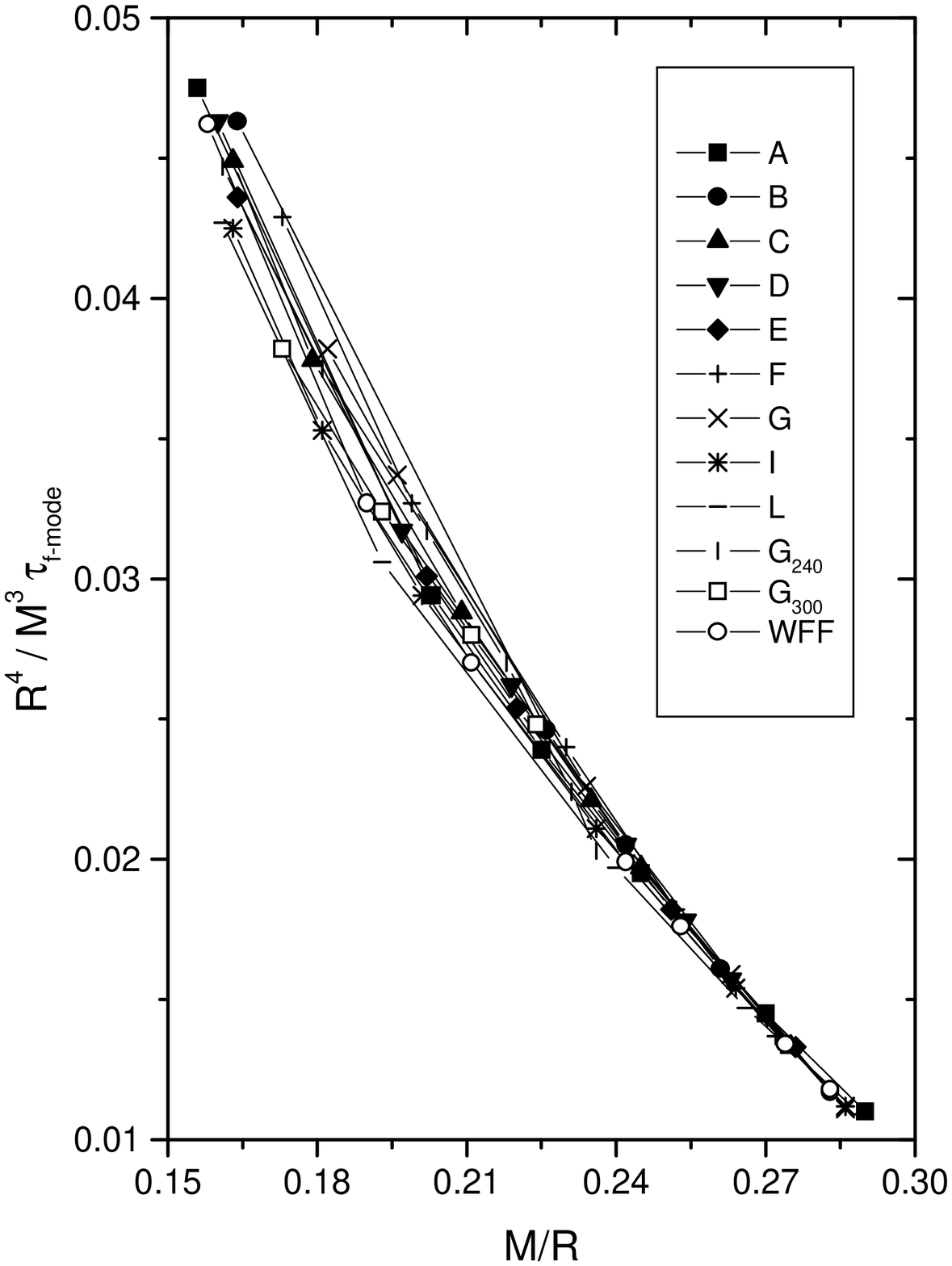}
\caption{Suitably normalized f-mode frequencies and damping rates 
for non-rotating stars constructed from  
a collection of realistic equations of state, cf. \cite{akeos}
for details.}
\label{moderes}\end{figure}

The  f-mode,  which can be viewed as the fundamental  
pressure mode of the star, is present already in a non-rotating star.
It corresponds to polar perturbations and, as was first shown by Kelvin, 
for a non-rotating uniform density star its frequency is 
given by
\begin{equation}
\omega^2_r = { 2l (l-1) \over 2l+1} { GM \over R^3}  \approx 
1.5\times10^8 { 2l (l-1) \over 2l+1} M_{1.4} R_{10}^3 \mbox{ s}^{-2}
\label{fmode}\end{equation}
This is a reasonable approximation also for more realistic equations 
of state, and we can readily deduce that a typical neutron star has 
f-mode frequency ($f=\omega_r/2\pi$) in the range 2--4~kHz (for $l=2$).
Mode frequencies and gravitational-wave damping rates for non-rotating stars 
constructed for many realistic equations of state are illustrated
in Figure~\ref{moderes}, cf. \cite{akeos}. 

Because of the symmetry of the non-rotating problem, modes 
corresponding to different
$l$ and $m$ decouple. In fact, it is sufficient
to consider the $m=0$ case. Rotation complicates the problem 
considerably.  First of all, it breaks the 
symmetry in such a way that the various $-l\le m\le l$ 
modes become distinct. As a first approximation one finds that
the f-mode frequencies change as
\begin{equation}
\omega_i(\Omega) =  \omega_r(\Omega=0)  + C_{lm} (\Omega) - m\Omega 
+ O(\Omega^2)
\label{rotfreq}\end{equation}
according to an inertial observer. Here, $\omega_r(\Omega=0)$ is the 
frequency of the mode in the non-rotating case, cf. (\ref{fmode}), 
and  $C_{lm}$ is a 
function that depends on the mode-eigenfunction in a non-rotating star.
Rotation also couples the various multipoles, which means that 
an increasing number of  $Y_{l}^m$'s
are needed to  describe a mode as the rotation rate
is increased. One must also account for coupling
between the polar and axial vectors. The problem is 
further complicated 
by the rotationally induced change in shape of the star, 
which first contributes at $O(\Omega^2)$ in the slow-rotation expansion.  

\begin{figure}[h]
\centering
\includegraphics[height=6cm,clip]{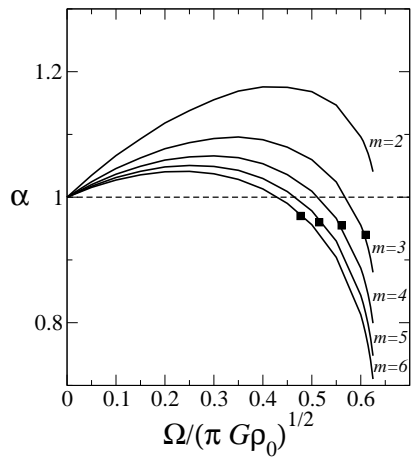}
\caption{This figure shows the  rotational corrections to the 
frequencies of the first few $l=m$ f-modes for a  
rotating Newtonian $n=1$ polytrope.
We graph  $\alpha = (\omega_i +m\Omega)/\omega_r(0)$
as a function of the rotation rate.  
Also shown, as filled squares, are the points at which these modes, which
are retrograde in the $\Omega=0$ limit, become prograde. [Data provided by Lee Lindblom.] }
\label{ilfig}\end{figure}

As we will see later, it is often useful to consider the ``pattern speed'' 
of the mode. Every mode of an axially symmetric system can be 
assumed to be  proportional
to $e^{i(m\varphi+\omega t)}$, and
surfaces of constant phase are therefore given by
\begin{equation}
m\varphi+\omega t = \mbox{ constant}
\end{equation}
After differentiation this leads to
\begin{equation}
{d\varphi \over dt}= -{\omega \over m} = \sigma
\label{pattern}\end{equation}
which defines the pattern speed $\sigma$ of the mode.
Having defined this quantity it is worth making
 two observations concerning the ($l=m$) f-modes.
1) From (\ref{fmode}) we see that the frequency of these modes
increases with $m$ roughly as $\omega_r \sim \sqrt{m}$. 
According to (\ref{pattern}) this means
that  the pattern speed of the f-modes  
decreases as we increase $m$. As a consequence one can always find an f-mode 
with arbitrarily  small pattern speed (corresponding to a  suitably 
large value of $m$) even though the high order f-modes
have increasingly large frequencies. 2) We can also see that mode 
patterns corresponding to opposite signs of $m$ tend to
rotate around the star in different directions. Taking the
positive direction to be that associated with $\Omega$ we 
find that the $l=\pm m$ modes are
backwards and forwards moving (retro/prograde), 
respectively, in the limit of vanishing rotation. 
However, rotation may change the situation for the $l=m$ f-modes. 
 (The particular case of the $m=\pm 2$ modes of a Maclaurin spheroid
is illustrated in Figure~\ref{macmodes}.)
Combining (\ref{fmode}) 
with (\ref{rotfreq}) --- where we neglect $C_{lm}$ for simplicity --- 
we can estimate that
these modes becomes prograde for rotation rates above
\begin{equation}
\Omega_s \approx \sqrt{ { 3 \over m} } \Omega_K  \quad \mbox{ or } 
\quad \beta_s \approx { 1 \over 3m}
\ . 
\label{prog}\end{equation} 
In other words, all but the $l=m=2$
f-modes are likely to change from backwards to forwards moving (according to 
an inertial observer) at realistic
rates of rotation ($\beta<\beta_K$).
This result is brought out clearly by more accurate calculations. 
Using their two-potential formalism, Ipser and Lindblom \cite{il1,il2}
have calculated the $l=m$ f-modes for rapidly rotating 
Newtonian polytropes. Their results are reproduced in 
Figure~\ref{ilfig}.
Similar results have been obtained by Yoshida and Eriguchi
\cite{ye1}. The numerical calculations show that the $m=3$ mode changes from 
retro- to prograde motion at $\beta \approx 0.08$ while the 
critical value for the $m=4$ mode is $\beta \approx 0.06$.

A non-rotating fluid star has no non-trivial axial modes (in Newtonian
theory). This situation is altered by rotation, which
leads to the presence of
inertial modes whose dynamics is governed by the Coriolis
force (see for example \cite{lfri} for a detailed 
discussion). A general 
inertial mode is such that $\delta \vec{v}$
is  composed of a mixture of polar and axial components
to leading order, i.e. one can take 
$[W_l, V_l,U_l] \sim \Omega$. 
With this ordering, all inertial modes are such 
that $[\delta p, \delta \rho] \sim \Omega^2$. 
The r-modes are a special subclass of inertial modes that 
(at least in Newtonian theory) are purely axial to leading order, i.e.
they have  $U_l \sim \Omega$ while $ [W_l,V_l] \sim \Omega^2$.
The r-mode frequencies are easy to determine from the fact that the 
radial component of the vorticity is conserved by their fluid motion,
and one finds that \cite{pap}
\begin{equation}
\omega_r \approx  {2m\Omega \over l(l+1)} \ .
\end{equation}

A non-barotropic star
has an infinite number of r-modes for each $l$ 
and $m\neq 0$ \cite{provost}.
In contrast, only a  single r-mode (for each $l=m$) 
exists in a barotropic star.
This mode is associated with the simple
fact that the star is a sphere and not a cylinder
(which is relevant since the Coriolis operator has 
cylindrical symmetry).
As a result, this mode is only weakly dependent on  internal 
stratification. In the general case the  radial dependency
of the r-mode eigenfunctions  remains undetermined at 
order $\Omega$, and the 
calculation must be taken to order $\Omega^2$.
The barotropic case is, again, different in that
one can determine that $U_l \propto r^{l+1}$ 
already at leading order \cite{provost}. 

The pattern speed for a typical $l=m$ r-mode is
\begin{equation}
\sigma_r = - {2 \Omega \over l(l+1) } 
\label{rotpat}\end{equation}
according to an observer rotating with the star.  
Meanwhile, an inertial observer would find 
\begin{equation}
\sigma_i = \Omega { (l-1)(l+2) \over l(l+1) }  \ .
\label{inpat}\end{equation}
That is, although the modes appear retrograde in the rotating system
an inertial observer would view them as prograde at all rotation rates. 

Given the eigenfunctions for a pulsation mode of a Newtonian star, 
we can use the multipole formulas in Section~\ref{Sec:gw} 
to estimate the rate at which the oscillation is damped by gravitational
radiation emission. As far as the f-modes are concerned, they are 
associated with significant density variations and one can easily show that 
the main contribution to the gravitational-wave damping comes from the 
mass multipoles. Detweiler \cite{det75}
has shown that, for uniform density stars, 
the gravitational-wave damping timescale can be estimated
as
\begin{equation}
t_{\rm gw} \approx { 2 \over 3} { (l-1)[(2l+1)!!]^2 \over (l+1)(l+2)}
\left[ {2l+1 \over 2l(l-1)}\right]^l \left({ c^2 R \over GM} 
\right)^{l+1} { R \over c}
\label{tgwnon}\end{equation}
or $t_{\rm gw} \approx0.07 M_{1.4}^{-3} R_{10}^4$~s for the quadrupole modes.
From this we see that the typical damping rate for the quadrupole
f-mode will be of the order of a tenth of a second.
Detailed results for a collection of realistic equations of state 
are given in Figure~2  of \cite{akeos}. 
 
An order count based on the multipole 
formulas suggests that the r-modes are different. 
For $l=m$ r-modes the dominant 
contribution to the gravitational radiation comes 
from the first term in the current multipole formula
(\ref{curr}). That this is the case 
can be seen as follows. The $l=m$ modes
have axial displacement to leading order $\sim \Omega$, while 
the density variation $\delta \rho$ 
enters at order $\Omega^2$.
Furthermore, we know that if the axial component
corresponds to the $l$th multipole, then the polar
components that arise from rotational coupling will correspond to $l+1$. 
This means that  we have $\delta D_{l+1m}\sim\Omega^2$ so $\dot{E}_{mass} \sim
\Omega^{2l+8}$,  while
$\delta J_{lm}\sim \Omega$  leads to $\dot{E}_{current} \sim \Omega^{2l+4}$. 
As a  gravitational-wave source 
the r-modes are therefore quite unusual.  The gravitational 
radiation that they emit comes primarily from the time-dependent 
mass currents, and is the gravitational analogue of 
magnetic multipole radiation. 

\subsection{Relativistic pulsations}
\label{grmodes}
 
Schematically, the relativistic problem consists of 
the perturbed Einstein equations
\begin{equation}
\delta G_{\mu \nu} = 8 \pi \delta T_{\mu \nu}
\end{equation}
and the equations of motion
\begin{equation}
\delta (T^\mu_{\nu;\mu} ) = 0 
\end{equation}
We need to solve (a subset of) these
equations for the four-velocity associated with the mode 
and the perturbed spacetime metric $\delta g_{\mu \nu}$.
In this calculation, the fluid four velocity is decomposed in 
a way that is completely analogous to the Newtonian 
description, cf. (\ref{vdec}), while the metric 
perturbations are represented  by the corresponding 
tensor decomposition (see \cite{thorne}). 
Just like in the Newtonian case, 
we can study axial and polar perturbations separately.
The relativistic perturbation equations have been derived many times over, 
and explicit formulas are given in (for example) \cite{relpert}.
A nice gauge-invariant description can be found in \cite{gund}.
The solutions that represent a pulsation mode satisfy the appropriate
regularity conditions at the centre of the star 
and correspond to purely outgoing waves at infinity. 
As already mentioned, the imposition of this boundary 
condition presents a technical challenge \cite{qnm}.
Several reliable techniques have been developed to handle this difficulty, 
and the pulsations of non-rotating relativistic stars
are by now well understood \cite{akeos}.
The spinning star problem still presents a challenge, however, and
the problem of calculating oscillation modes of rapidly rotating
neutron stars in General Relativity (including the imaginary 
parts of $\omega$ associated with gravitational-wave damping)
remains unsolved.

As we will see later, the neutral modes (representing the point where 
an originally retrograde mode becomes prograde)
of a rotating star determine the onset of gravitational-wave
driven instability. Consequently, one can focus attention on 
the simplified problem of finding time-independent 
solutions to the perturbation equations. Such solutions would be marginally unstable.
A numerical solution to this problem was obtained by Stergioulas
and Friedman \cite{sf} for polytropic stars. Their results show that 
relativistic effects tend to destabilize a rotating star
considerably. 
Interestingly, one finds that in  General Relativity
the $m=2$ f-mode may  have a neutral
point for attainable rates of rotation. This  
result, which constrasts with the Newtonian case shown in Figure~\ref{ilfig}, 
has been shown to hold also for realistic equations
of state \cite{msb98}. 
An empirical fit to results for several realistic equations
of state suggests that the $m=2$ f-mode becomes secularly unstable
at 
\begin{equation}
\beta_s \approx 0.115 - 0.048{ M \over M_{\rm max}(\Omega=0)}
\end{equation}
where $M_{\rm max}(\Omega=0)$ is the maximum allowed mass of 
a nonrotating star for the given equation of state.
For a typical
$1.4M_\odot$ star the $m=2$ f-mode has a neutral point
near $\beta \approx 0.08$ or $\Omega \approx 0.85\Omega_K$.    
This result could be of considerable importance since 
mass quadrupole radiation is likely to lead to the fastest 
instability growth.
One further step towards the calculation of f-modes of 
rotating relativistic stars was taken by Yoshida
and Eriguchi \cite{ye2,ye3}. They solved the problem 
in the relativistic Cowling approximation, wherein one assumes
that $\delta g_{\mu \nu} = 0$. The obtained results
show that the Cowling approximation tends to 
overestimate the stability of the star, and therefore
emphasize the conclusion that relativity has a destabilizing
effect. 

General Relativity also affects the r-modes in a significant
way. This is perhaps not  surprising since the
relativistic framedragging is an order $\Omega$ effect, 
which may affect inertial modes at ``leading order''. 
Retaining only the leading order rotational effects, 
the metric of a stationary equilibrium can be written
\cite{hartle}
\begin{equation}
ds^2 = -e^{2\nu(r)} dt^2 + e^{2\lambda(r)} dr^2 + r^2 d \theta^2 
+ 
r^2 \sin^2 \theta d \varphi^2  - 2 \omega(r) r^2 \sin^2\theta dt d\varphi
\end{equation}
The frame-dragging is represented by the function $\omega(r)$.
(To avoid confusion with the mode-frequency $\omega$ we will always
retain the dependence on $r$ in expressions involving the 
frame dragging.)

Although some details remain 
to be understood, significant progress has been made
on the  relativistic r-mode problem in the 
last couple of years. 
For barotropic relativistic stars one can prove that 
no purely axial inertial modes can exist \cite{letal}.
All inertial modes of such stars 
are a hybrid mixture of axial and polar perturbations
to leading order. For the particular modes that limit to 
the purely axial r-modes as $M/R \to 0$ one can show 
(for uniform density stars) that 
\cite{letal}
\begin{equation}
\kappa = \frac{2}{(m+1)}\left[1
-\frac{4(m-1)(2m+11)}{5(2m+1)(2m+5)}\left(\tmr\right)
+ O\left(\tmr\right)^2 \right]
\label{kap}\end{equation}
where $\kappa = (\omega+m\Omega)/\Omega$.
Meanwhile, the velocity eigenfunctions take the form
\begin{equation}
U_m(r) = \left(\rx\right)^{m+1}\left[
1+C\left(1-\frac{r^2}{R^2}\right)\left(\tmr\right)
+ O\left(\tmr\right)^2\right]
\end{equation}
where $C$ is a constant, and 
\begin{equation}
W_{m+1}\sim 
V_{m+1} \sim U_{m+2} \sim  O\left(\tmr\right) 
\end{equation}
From this we  see how the $l=m$ Newtonian r-modes of 
barotropic uniform density stars are  affected
by first order post-Newtonian corrections.
All barotropic Newtonian 
r-modes with $m\geq 2$ pick up both axial and polar
post-Newtonian terms. From Eq. (\ref{kap}) 
we also see that the r-mode frequency decreases because of the small relativistic 
correction. It is natural that 
General Relativity will have this effect. The
gravitational redshift will tend to decrease the fluid oscillation 
frequency as measured by a distant inertial observer. Also, because
these modes are rotationally restored they will be affected by the
dragging of inertial frames induced by the star's rotation.
Specifically, since the Coriolis force is determined by
the fluids 
angular velocity relative to that of the local inertial frame
\cite{hatho},
$\bar\om(r)= \Omega -\omega(r)$ it decreases --- 
and the modes oscillate less rapidly --- as the dragging of inertial 
frames becomes more pronounced.  

\begin{figure}[h]
\centering
\includegraphics[height=6cm,clip]{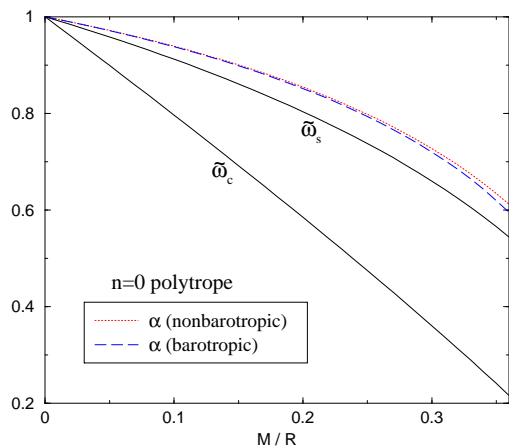}
\caption{This figure shows the 
r-mode eigenfrequencies $\alpha$ for relativistic nonbarotropic 
uniform density stars ($n=0$ polytrope) and a range of compactness ratios
$M/R$. The Newtonian limit corresponds to $M/R\to 0$.
 Also shown 
are the corresponding values of the relativistic framedragging at the centre 
$\tilde{\omega}_c$ and surface $\tilde{\omega}_s$ of the star. The 
perturbation equations admit a continuous spectrum in the range 
$\tilde{\omega}_c < \alpha < \tilde{\omega}_s$. For such frequencies
 the eigenvalue problem is formally singular. As is clear from the data, 
the uniform density r-modes are
always regular. However, it should be noted that most 
realistic equations of state
lead to a singular problem.
Also shown (as a dashed curve) are the eigenfrequencies for the 
axial-led inertial mode of a barotropic star that most resembles the 
Newtonian r-mode.  Note that the hybrid/inertial 
mode problem is never
singular.}
\label{relr}\end{figure}

Just like in  Newtonian theory, the non-barotropic problem is
different. For non-barotropic stars one can 
still have purely axial inertial modes also in General Relativity. 
These modes are 
determined by a single ordinary differential equation
for one of the perturbed metric components;
\begin{eqnarray}
(\alpha -\tilde{\omega})
\left\{  e^{\nu-\lambda} {d\over dr} \left[ e^{-\nu-\lambda} 
{dh \over dr} \right] \right. &-& \left.\left[{l(l+1) \over r^2} - {4M\over r^3}
+8\pi(p+\rho) \right]  h \right\}  \nonumber \\ 
&+& 16\pi(p+\rho) \alpha h = 0 \ ,
\label{singeq2}\end{eqnarray}
where 
 $h \propto \delta g_{t\varphi}$, and 
we have used 
\begin{equation}
\omega = -m\Omega\left[ 1 - {2\alpha\over l(l+1)} \right] \ ,
\end{equation}
as well as $\tilde{\omega}= \bar{\omega}/\Omega$.
This equation was first derived by Kojima \cite{kojima}. 
The eigenvalues $\alpha$ and the corresponding 
eigenfunctions $h$ are not explicitly dependent on $m$, which
 means that if we find an acceptable mode-solution
to (\ref{singeq2}) it will be relevant for all $m\neq 0$ for each given 
multipole $l$. This would be in accord with the non-barotropic
Newtonian case where one finds a single r-mode for each combination of 
$l$ and $m$ at order $\Omega$. Given (\ref{singeq2}) one can prove
two interesting results. First of all, 
non-trivial solutions  may only exist  provided that 
 $\alpha-\tilde{\omega}$  vanishes at
some point in the interval $r\in [0,\infty]$ \cite{letal}.
As first demonstrated  by Lockitch et al 
\cite{letal} for uniform density stars, 
one can find a single discrete mode solution
to (\ref{singeq2}) with frequency in the required interval, cf. Figure~\ref{relr}. 
Secondly, the equation admits a continuous spectrum \cite{kojima,kb} 
in the range $\tilde{\omega}(0) < \alpha < 
\tilde{\omega}(R)$. The dynamical role of this continuous spectrum, 
or indeed if it remains present when higher order rotational corrections
are included, is not clear at the present time. 
The presence of the continuous spectrum makes the non-barotropic 
r-mode problem difficult. When one considers softer equations of state
one must typically  determine a discrete mode lying inside 
the continuous spectrum. As is well known from studies of 
differentially rotating systems (where a continuous spectrum 
arises because of the presence of so-called co-rotation points)
this is a notoriously difficult problem \cite{balbinski,anna}. 
This technical difficulty has led to suggestions that 
the r-modes may not even ``exist'' for certain relativistic
stars \cite{yosh,kr01}. However, such conclusions are likely
premature: A failure to calculate the mode is more likely 
an indication of a breakdown in the approximations 
we have made to the physics. In the case of the r-modes it is clear
already from (\ref{singeq2}) that the 
slow-rotation approximation is no longer
consistent in regions where $\alpha - \tilde{\omega} \sim O(\Omega^2)$ 
or smaller.
In principle, this means that the problem requires a ``boundary 
layer'' approach \cite{regul} where either $\Omega^2$ terms, 
viscosity or the coupling to polar perturbations are included
in the analysis. Some very recent results \cite{yl02,rsk02}
support this view. 

Finally, it worth mentioning that there is a class of pulsation 
modes that arise only when the pulsation problem is 
considered in General Relativity.
These are known as the w-modes \cite{ks}, and they exist because the curvature
of spacetime that is generated by the background density distribution
can temporarily trap impinging gravitational waves. The w-modes typically
have high frequencies (above 7~kHz) and they damp out in a fraction of a 
millisecond (see data provided in \cite{akeos}). 
For ultracompact stars these modes may become very long lived
\cite{trapped}. 
Gravitational waves can be trapped inside the peak of
the spacetime curvature
barrier that is unveiled as $R<3M$ (in the non-rotating case).
Even though such extremely compact stars may not exist in 
the Universe (no proposed realistic equations of state permit 
stars more compact than $R\approx 3M$),  
they are still intriguing. In particular since, 
when rotating, they may admit the presence of an ergosphere, 
i.e. a region (located inside the star) in which all observers must be dragged 
along with the stars rotation. This could lead to \cite{ergo1}
the w-modes
becoming unstable due to the same mechanism that drives
the CFS unstable fluid modes (see Section~\ref{Sec:sec}).
This instability is unlikely to be astrophysically relevant \cite{ergo2,ergo3,ergo4}, 
but since it is a uniquely relativistic effect it is
of conceptual interest. 

\section{What do we learn from the ellipsoids?} 
\label{el_sec}

Many aspects of the instabilities in spinning stars
can be illustrated by classic results
for rotating ellipsoids. This is, in fact, an age old
problem that has attracted the attention of distinguished scientists
for  centuries\footnote{The subject has been exhaustively summarized
by Chandrasekhar \cite{efe}.}. The relative mathematical simplicity
makes a study of the equilibrium properties and stability 
of rotating self-gravitating fluid bodies with uniform density
analytically tractable. In this Section we will discuss the key results
and try to make contact with the pulsation properties
of compressible stars. 

The most studied (and therefore best understood) figures of 
 rotating, self-gravitating, homogeneous and incompressible
fluid bodies in equilibrium are the uniformly rotating Maclaurin spheroids, which
are oblate in shape. In addition, there exist various triaxial equilibrium 
configurations. The Jacobi ellipsoids are rigidly rotating about the smallest axis,
and have no vorticity when viewed from a rotating frame in which the 
figure appears stationary. 
For a given angular momentum, mass and volume the
Jacobi ellipsoid has lower energy than the corresponding Maclaurin
spheroid. The Dedekind ellipsoids have a stationary triaxial shape in the inertial frame. 
Hence, they are non-rotating, and their shape is entirely supported 
by internal motions of uniform vorticity. A Dedekind
ellipsoid with the same mass and circulation as the corresponding
Maclaurin configuration has lower angular momentum. 
 These three families of equilibrium spheroids are 
subclasses of the so-called Riemann-S ellipsoids, which 
are distinguished by having  rotation and vorticity vectors 
aligned with a symmetry axis of the figure. The Riemann sequences
are characterized by having constant ratio $\zeta/\Omega$, where
$\zeta$ is the vorticity in the rotating frame. The 
Jacobi and Dedekind families correspond to the special
cases $\zeta=0$ and $\Omega=0$, respectively. 

\begin{figure}[h]
\centering
\includegraphics[height=6cm,clip]{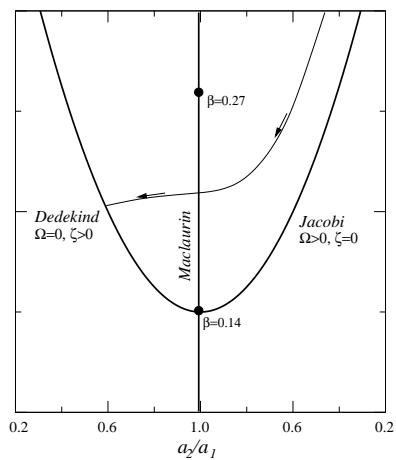}
\caption{A schematic summary of the instability results
for rotating ellipsoids ($a_2/a_1$ represents the axis ratio, 
i.e. the ellipticity of the configuration). For values of $\beta$ greater than 
0.14 the Maclaurin spheroids are secularly unstable. Viscosity 
tends to drive  the system towards a triaxial Jacobi 
ellipsoid, while gravitational radiation leads to an evolution
towards a Dedekind configuration. Indicated in the figure is an
evolution of this latter kind. Above $\beta \approx 0.27$ the 
Maclaurin spheroids are dynamically unstable, as there exists a
Riemann-S ellipsoid with lower (free) energy.
[For more details, see \cite{lais,christo}]}
\label{ellipse}\end{figure}

We being by focussing our attention on the Maclaurin spheroids.
Proceeding along the Maclaurin sequence towards more rapidly 
rotating configurations, eg. increasing $\beta$, one finds    
a bifurcation point at $\beta_s \approx 0.14$. At this point
 the  Jacobi and Dedekind ellipsoids both branch off
from the Maclaurin sequence.
This is schematically illustrated in Figure~\ref{ellipse}.
Given the existence of the alternative states
with lower energy/angular momentum beyond 
the point of bifurcation it would be
favourable for a perturbed Maclaurin spheroid to
move towards either the Jacobi or the Dedekind sequence.
However, this is not possible as long as 
the system conserves circulation and angular momentum.
This means that   
the Maclaurin spheroid will remain stable unless we add
dissipation to the dynamical equations. In other words,
the bifurcation point at $\beta_s$ indicates the onset
of secular instabilities. 

Viscosity dissipates energy while preserving the angular momentum.
The Maclaurin spheroids are therefore  susceptible to a 
viscosity-driven instability once $\beta > \beta_s$, and 
the instability drives the system towards the Jacobi 
sequence \cite{roberts}. Gravitational-wave dissipation, on the other hand, 
radiates angular momentum while conserving the internal circulation.
Thus, the Maclaurin spheroids also suffer a 
gravitational-wave driven instability when $\beta > \beta_s$.
The gravitational-wave instability tends to drive the system 
towards  the Dedekind sequence (the members of which 
do not radiate gravitationally)\footnote{Recent
results concerning the stability of the Riemann-S ellipsods  complicates 
this picture considerably. These results, due to Lebovitz and 
Lifschitz \cite{ll96}, show that the Riemann-S ellipsoids suffer a
``strain'' instability in most of the parameter space. 
In particular, the Dedekind ellipsoids are always unstable due to this new 
instability.}.

\begin{figure}[h]
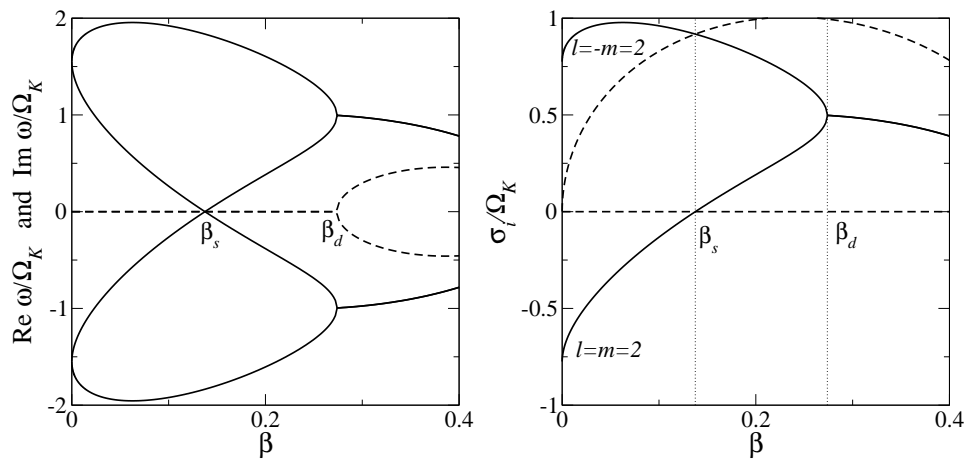

\centering
\includegraphics[height=6cm,clip]{macmodes2.eps}
\includegraphics[height=6cm,clip]{macmodes.eps}
\caption{Results for the $l=|m|=2$ f-modes of a Maclaurin spheroid. In the 
left frame we show the oscillation frequencies (solid lines) and imaginary parts 
(dashed lines) of the modes, while the 
right frame shows the mode pattern speed $\sigma_i$
for the two modes that have positive frequency 
in the non-rotating limit (the pattern speeds for the modes which have negative 
frequency in the non-rotating limit
are obtained by reversing the sign of $m$). 
All results are according to an observer in the inertial frame.
The dashed curves in the right frame 
represent a vanishing pattern speed i)  in 
the inertial frame (the horizontal line), and ii) in the rotating frame 
(the circular arc, which 
shows $\Omega/\Omega_K$ as a function of $\beta$). 
The points where the 
Maclaurin ellipsoid becomes secularly ($\beta_s$) and dynamically ($\beta_d$) 
are indicated by 
vertical dotted lines.}
\label{macmodes}\end{figure}

These classic secular instabilities set in through the quadrupole
f-modes of the ellipsoids. In Figure~\ref{macmodes}
we show the frequencies of the $l=|m|=2$ Maclaurin spheroid f-modes.
These modes are usually refered to as the ``bar-modes''.
The figure illustrates several general features of the pulsation 
problem for rotating stars. In particular we notice i) the rotational splitting
of modes that are degenerate in the non-rotating limit, i.e. the $m=\pm2$ modes
become distinct in the rotating case, and ii) the symmetry with respect to
$\omega=0$, which reflects the fact that the governing equations are invariant under the 
change $[\omega,m]\rightarrow[-\omega,-m]$.
In Figure~\ref{macmodes} we also show the pattern speed for the two modes 
that have positive frequency 
in the non-rotating limit, cf. (\ref{fmode}).
 From this figure we see that the $l=-m=2$ mode, which is always prograde
moving in the inertial frame, has zero pattern speed in the rotating frame 
at $\beta_s$ ($\sigma_p =\Omega$). At this point the mode becomes unstable to the 
viscosity driven instability. That the instability should set in at this point is natural
since the perturbed configuration is ``Jacobi-like'' when 
the mode is stationary in the rotating frame. 
Meanwhile, the  gravitational-wave instability sets in 
through the originally retrograde moving $l=m=2$ modes. At $\beta_s$ these
modes have zero pattern speed in the inertial frame ($\sigma_p=0$).
At this point the perturbed configuration is ``Dedekind-like'' since 
the mode is stationary according to an inertial observer.

The evolution of the  secular instabilities
depends on the relative strength of the dissipation mechanisms. 
This tug-of-war is typical of these kinds of problems.
Since the gravitational-wave driven mode involves
differential rotation it is damped by viscosity, and since the
viscosity-driven mode is triaxial it tends to be damped by 
gravitational-wave emission. A detailed understanding of the
dissipation mechanisms is therefore crucial for any investigation into 
secular instabilities of spinning stars. 

Given the competition between gravitational radiation and viscosity
one would expect a ``realistic'' star to be stabilized beyond
the point $\beta_s$. Also, the secular instabilities are no longer realized 
in the extreme
case of a perfect fluid which conserves both angular momentum 
and circulation\footnote{Note
that in General Relativity all non-axisymmetric modes of oscillation radiate
gravitational waves. Hence, this argument is only relevant in Newtonian gravity.}. 
Then the Maclaurin sequence
remains stable up to the point $\beta_d \approx 0.27$.  
At this point there exists a bifurcation to the $x=+1$ Riemann-S sequence. 
These equilibria have lower ``free energy'' \cite{christo} than the corresponding 
Maclaurin spheroid for the same angular momentum and circulation. 
This means that a dynamical transition to a lower energy state
may take place without violating any conservation laws. In other words, 
at $\beta_d$ the Maclaurin spheroids become dynamically unstable to  $m=2$
perturbations. This instability is usually refered to as the dynamical bar-mode 
instability. 

In terms of the pulsation modes, the dynamical instability
sets in at a point where two real-frequency modes merge, cf. Figure~\ref{macmodes}.
At the bifurcation point $\beta_d$ the two modes have identical oscillation
frequencies and their angular momenta will vanish. 
Given this, one of the degenerate modes can  grow without
violating the conservation of angular momentum. The physical conditions 
required for the dynamical instability are easily understood. 
The instability occurs when the originally backwards
moving f-mode (which has $\delta J<0$ for $\beta<\beta_d$)
has been dragged forwards by rotation so much that 
it has ``caught up'' with the originally forwards moving mode
(which has $\delta J>0$ for $\beta>\beta_d$).
In order for the modes to merge and become degenerate
the perturbation must have vanishing angular momentum  at $\beta_d$ 
($\delta J= 0$).

\section{Stability analysis}

Stellar stability problems have traditionally been 
explored either via construction of suitable variational
principles or  direct mode-calculations. 
Each strategy has its  merits and drawbacks. 
The variational-principle approach is appealing from a formal point of view, 
and it is well known from many branches of physics that variational
principles are closely connected to stability criteria. 
Once one has defined a suitable ``energy'' for the perturbations, 
stability simply follows from its sign 
for any given perturbation. However, the obtained relations can be 
difficult to use in  practice. It may also be difficult
to conclusively rule out instabilities.  
The mode-calculation approach is less elegant, but  has the advantage that 
one can test each individual mode for stability. On the other hand, 
a failure to find an unstable mode does not necessarily 
mean that one does not exist. 

The main development of the 
theoretical framework for studying stellar stability in general 
relativity took place in the 1970s. 
Key early contributions were made
by Chandrasekhar and Friedman \cite{cf72a,cf72b} and Schutz 
\cite{bfs72a,bfs72b}. 
Their work attempted to extend the conclusions from 
Newtonian studies --- that the onset of non-axisymmetric 
instability is signalled by the appearance of a neutral (zero-frequency) mode.
There are two main reasons why a relativistic analysis
is significantly more complicated than the Newtonian one. 
First of all the problem is algebraically more complex
because one must solve the Einstein field equations
in addition to the fluid equations of motion. Secondly, 
one must account for the fact that a general 
perturbation 
will generate gravitational waves. This is a fundamental 
complication since a 
mode-based proof of stability would require some kind of completeness
of the modes of the star. Since the relativistic 
modes are ``quasinormal'' (they have complex frequencies, with the
imaginary part corresponding to the gravitational-wave damping)
they are unlikely to be complete in any meaningful sense \cite{qnm}.  

The work of Friedman and Schutz culminated in a series
of impressive papers  \cite{fs0,fs1,fs2} in which the role that gravitational 
radiation plays in these problems was explained, 
and a foundation for subsequent research in this area
was established. The main result was that gravitational radiation
acts in the same way in the full theory as in 
Chandrasekhar's post-Newtonian
analysis of the Maclaurin spheroids \cite{chandra70}.
If we consider a sequence of equilibrium models, then
a mode becomes secularly unstable at the point where
its frequency vanishes (in the inertial frame). Most importantly, the 
proof does not require the completeness of the modes of the system.  

The Friedman-Schutz criterion for instability
relies on the so-called canonical energy $E_c$ being negative.
The canonical energy is defined as
\begin{eqnarray}
E_c &=& {1\over 2} \int \left[ \rho |\partial_t \vec{\xi}|^2
-\rho|\vec{u}\cdot\nabla\vec{\xi}|^2 +\Gamma_1 p 
|\nabla \cdot \vec{\xi} |^2 + \vec{\xi}^*\cdot\nabla p
\nabla\cdot\vec{\xi} +  \right. \nonumber \\
&+& \left. \vec{\xi}\cdot\nabla p\nabla\cdot
\vec{\xi}^* + \xi^{i*} \xi^{j*}(\nabla_i\nabla_j p +
\rho\nabla_i\nabla_j \Phi) - { 1 \over 4 \pi G} | \nabla \delta \Phi |^2 
\right]  dV
\end{eqnarray}
where $\vec{u}= \Omega\times \hat{r}$ 
represents the background flow\footnote{A fully relativistic 
expression for the canonical energy has been derived 
by Friedman \cite{jf}.}. 
One can also define a (conserved) canonical angular 
momentum 
\begin{equation}
J_c = -\mbox{Re } \int \rho \partial_\varphi \xi^{i*}(
\partial_t {\xi}_i+\vec{u}\cdot\nabla\xi_i) dV
\end{equation}

The canonical energy and angular momentum are conserved in absence of 
radiation and viscosity. This means that, in order to have a dynamical
instability (unbounded growth of a linear mode of the inviscid problem)
we must have $E_c=J_c=0$.
If $E_c$ is negative at the outset and the star
is coupled
to radiation in such a way that  $E_c$
must decrease with time, then the absolute 
value of $E_c$ will increase and the associated mode
will be unstable. Generally, an instability can be established 
in a mode-independent way by constructing
(canonical) initial data $[\vec{\xi},\partial_t \vec{\xi}]$ such 
that $E_c$ is negative. 
To do this is, however, not a simple task.
From the computational point of view  it is easier to 
calculate a
mode of a rotating star and then evaluate $E_c$ to assess stability. 
It is sufficient to show that the displacement
vector associated with the mode leads to $E_c<0$ to 
demonstrate the presence of an instability.

The simple intuitive instability criterion 
can be deduced from the relation  
\begin{equation} 
E_c = -{\omega_i \over m} J_c = \sigma_i J_c
\label{ec}\end{equation}
which is a general property of linear waves. We see that $E_c$ 
changes sign when the inertial frame 
pattern speed $\sigma_i$ passes through zero.
Beyond this point the mode moves forwards with respect to the
inertial frame while it is  still moving 
backwards in the rotating frame. 
Gravitational waves from such a 
mode carry positive angular momentum 
away from the star, but since the perturbed fluid actually 
rotates slower than it would in  
absence of the perturbation the angular momentum of the 
mode is negative. The emission of 
gravitational waves consequently
makes the angular momentum of the mode 
increasingly negative and leads to the instability. 
This  instability is  often
refered to as the Chandrasekhar-Friedman-Schutz
(CFS) instability.

It is interesting to contrast the secular radiation
driven instability to that associated with viscosity. 
For uniformly rotating stars one can show that 
the combination
\begin{equation}
\delta E - \Omega \delta J = E_c - \Omega J_c = -
{\omega_r \over m} J_c =\sigma_r J_c = E_{c,R}
\label{ecrot}\end{equation}
relating the first order changes in the kinetic
energy and angular momentum to a mode-solution, is gauge-invariant. 
 $E_{c,R}$ can be viewed as the canonical energy in 
the rotating reference frame. 
Viscosity leads to $E_{c,R}$ being a decreasing
function of time. From (\ref{ecrot})
we can deduce that the onset of the viscosity driven instability is 
signalled by the vanishing of the mode pattern speed in the
rotating frame ($\sigma_r=0$).

In order to investigate whether an instability is 
of astrophysical relevance, eg. whether it leads to a detectable
gravitational-wave signal, one must address two main
questions. First of all, one must understand under what 
circumstances the instability will be present and how likely
it is that a star will evolve through the relevant 
part of parameter space. Secondly, one must establish
that the unstable mode grows on a sufficiently 
short timescale. This is always the case for 
dynamical instabilities, but as we will see in the following sections 
the issues that decide whether a secular instability
is relevant or not are much more delicate.

\section{Results: Dynamical instabilities}

\subsection{Quasiradial modes}

The most familiar stellar instability is probably that
associated with the 
existence of a maximum mass configuration for any given 
equation of state. A spherical 
relativistic star will 
suffer a dynamical instability before the compactness
reaches the Schwarzschild limit \cite{chandra64}.
In General Relativity you can
never have stable stars with $R<2.25M$ and no realistic
equations of state  permit stars more compact than
$R\approx 3M$.
Once an accreting neutron star reaches the maximum mass 
limit it will become unstable and undergo gravitational collapse, 
most likely leading to the formation of a black hole.   

The maximum mass limit provides one of very few handles that we
currently have on the supranuclear equation of state. Several
neutron star masses have been deduced from pulsar observations
and, in order to be acceptable, a proposed equation of state 
must allow for masses at least as large those observed. 
The data for the binary pulsar PSR1913+16  provides the constraint 
$M_{max}\ge 1.44M_\odot$, which allows us to 
rule out extremely soft equations of state. 
A much more
severe constraint on the theoretical models may be provided by the data 
for the Vela pulsar. Several studies have estimated the 
mass of Vela to be about $1.8M_\odot$ \cite{velamass}.
Many proposed equations of state would be in trouble if this 
result were to be
confirmed. In particular, neutron star models with sizeable
exotic cores composed of hyperons and/or deconfined quarks 
which tend to soften the equation of state
\cite{glend} might then be ruled out.

The maximum mass instability is relatively easy to analyse in the case of 
non-rotating stars. It sets in through the radial ($l=m=0$) f-modes.
Since the equations that describe radial oscillations depend only
on $\omega^2$ the mode frequencies come in pairs ($\pm \omega$). 
Calculations show that, as one increases the central density of the star
(for a given equation of state) the absolute value of
the stable (real valued) f-mode frequencies 
decreases. It passes through zero at the point at which the 
mass reaches an extremum. Beyond that ``turning point'', the mode frequencies
become a complex conjugate pair, and thus one of the modes
is unstable.

\begin{figure}[h]
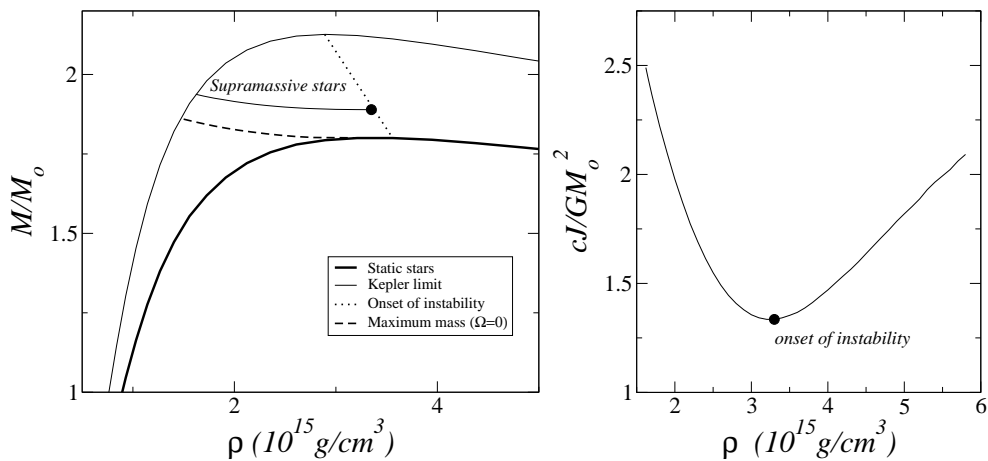

\centering
\includegraphics[height=6cm,clip]{FPS.eps}
\includegraphics[height=6cm,clip]{min.eps}
\caption{Left panel: The permissible region of stable stars described 
by the FPS equation of state. Rotation increases the maximum mass
by roughly 20\% and also leads to the presence of a family
of stars that have no non-rotating counterpart.
These configurations are located between the mass-shedding curve
and the dashed curve, which represents the most massive rotating star
that has a stable non-rotating counterpart.
A particular sequence (with constant baryon mass)
of such supramassive stars
in indicated by a thin solid line. An isolated star that spins down due to 
magnetic dipole radiation would evolve along this sequence until
it reaches the state represented by the filled circle. At this point 
it will become unstable 
and undergo gravitational collapse.
Right panel: The instability of a supramassive
star sequence (with constant baryon mass)
sets in at the point where the angular momentum $J$ 
has an extremum as a function of the central density.  
}
\label{FPS}\end{figure}

The simplicity of the ``turning point method'' 
for locating the onset of instability along
a sequence of equilibrium models is extremely appealing.
It is, however, not immediately obvious that it will
generalize to a more complicated setting --- like rotating stars.  
The problem is, in fact, quite subtle.  A
dynamical stability analysis requires evaluation of 
a complicated energy functional, and then  
relies on non-degenerate perturbation theory to deduce
stability. However, 
instabilities set in through zero-frequency modes, and  
the simple non-rotating barotropic models that one would typically
consider in this context have  a degenerate set of 
such neutral modes --- the inertial modes. These would 
be distinct in a non-barotropic model
(in which they become the g-modes), but in 
the barotropic case they all have zero frequency. 
A proper stability analysis must take the presence
of these modes into account. In order to single out a ``marginally
unstable'' mode from the sea of inertial modes one must use degenerate
perturbation theory. This is not a simple task.
Fortunately, Sorkin \cite{sorkin} has provided a detailed proof that the
turning point method can be used 
to locate the onset of instability. The proof 
does not assume mode-completeness and hence neatly circumvents  
difficulties like that associated with the degenerate inertial modes.
In the case of uniform rotation an
extremum of the angular momentum $J$ along a sequence 
with constant baryon number limits the region of stable stars
\cite{fis88}. 
The (dynamically) stable region for a typical relativistic star 
(represented by the FPS equation of state) is illustrated in 
figure~\ref{FPS}. As can be seen in the figure, 
rotation generally increases the maximum allowed mass by up 
to 20\%. It is  worth noting that there may exist a class 
of rotating stars that have no non-rotating counterpart.
These ``supramassive'' stars must eventually collapse if they are 
spun down, eg. by magnetic dipole radiation \cite{cook}.

Sorkin's theorem states that for barotropic stars 
a turning point in $J$ marks the location of 
a zero-frequency axisymmetric mode and the onset of dynamical 
instability. For non-barotropic stars the situation is a bit more subtle.
In that case the turning point method locates a point where
a secular instability sets in. 
Consequently, a  
realistic star is likely to undergo a secular instability phase 
before it reaches the dynamical instability point and collapses.

From an intuitive point of view one might expect gravitational
collapse to lead to a very strong gravitational-wave
signal. However, it is also conceivable that the level of radiation 
may be  low. The outcome depends entirely on the asymmetry 
of the collapse process. A purely spherical collapse
will obviously not radiate gravitationally at all, while the
collapse of a strongly deformed body could release a copious 
amount of gravitational waves. The main reason why it is very difficult
to make ``reliable'' estimates for the energy released is
that the answer depends  entirely on the route that the system follows
towards the final configuration. 
This is immediately clear from the post-Newtonian
formulas (\ref{gwlum}), 
which show that the gravitational-wave luminosity depends
on high time-derivatives of the various multipoles.

It was originally thought that supernova core-collapse
would lead to strong gravitational-wave signals.
This expectation has not been confirmed by detailed simulations.
In fact, the available numerical simulations paint a somewhat 
pessimistic picture. Typical 
results suggest that an energy equivalent to $8\times 10^{-8} M_\odot c^2$ 
may be radiated \cite{zwmull}. The signal from a collapse that leads to the 
formation of a black hole is likely to be dominated by the 
slowest damped quadrupole quasinormal mode of a black hole, i.e.  have 
frequency 
\begin{equation}
f_{Hz} \approx 1200\left( {10 M_\odot \over M} \right) 
\end{equation}
If we assume that a typical timescale for the 
event is of the order of a millisecond we find, cf. (\ref{heff1}),
that the gravitational-wave amplitude may be of the order of
 $h_c  \sim  10^{-22}$
for a source in the Virgo cluster. 
This estimate (which accords reasonably well with full numerical 
simulations) suggests that such 
sources are unlikely to be observable beyond the local group 
of galaxies. 

Until very recently studies of core collapse focussed mainly on
the non-rotating case (the classic work by Stark and Piran 
\cite{sp} being a notable exception).
It is perhaps natural to expect that studies of
the fully three-dimensional collapse problem
may lead to an enhanced radiated energy
and maybe even a prediction of clearly detectable waves.
To solve  this problem is, of course, a far from
trivial task. Nevertheless, there has been interesting 
recent progress in this direction \cite{dimmel,shibata}. 
Numerical relativity is rapidly maturing
and it may not be too optimistic to expect that this problem 
will be manageable in the not too distant future. Having said that, it 
is clear that there are conceptual issues that may be very hard to 
resolve, in particular regarding the initial data.  
Numerical relativists have so far almost exclusively considered
initial data obtained using York's conformally flat prescription.
This approach is taken because of its mathematical convenience.
It is reasonable  to worry about this since there
is absolutely no physical reason why ``realistic'' 
initial data should be conformally flat. This is a key
issue that needs to be addressed in the future.  

In absence of detailed numerical results, 
it is interesting to speculate about the outcome of the 
gravitational collapse of a rapidly rotating star. 
It stands to reason that this would be a promising source
for gravitational radiation, since rotation will couple 
the various multipoles in such a way that even quasi-radial 
modes will radiate. Furthermore, it is well known that 
rapidly rotating stars have a multipolar structure that 
differs significantly from that of a Kerr black hole (cf. the 
results in \cite{lp99}).
In order to form a black hole the collapsing star must
in some way shed the ``difference'' in the various multipole moments.
Presumably, this will be done mainly through the emission 
gravitational waves. 
Of course, a sceptic would argue that the collapse
event may  proceed in such a way that the level of radiation is minimal. 
Future multidimensional simulations will have to provide the 
real answer. 

A collapse related
scenario that has been discussed  in the literature
concerns internal phase-transitions. As a neutron star
is spinning down, eg. due to magnetic dipole radiation, the 
 central density will increase.
Various theoretical models suggest that the equation of state
may soften significantly once the central density 
increases beyond a critical value (several times nuclear 
density). This could be due to the formation of 
pion/kaon condensates, the creation of a significant hyperon core
or quark deconfinement. Should this happen 
it is likely to result in  a ``mini-collapse'' during which 
some gravitational potential energy may be released as radiation. 
Such phase transitions have been suggested as sources for both 
detectable gravitational waves \cite{cd98} and gamma-ray bursts.
However, most of the available estimates seem 
somewhat optimistic. The reason 
for this is very simple. It is typically assumed that the entire 
change in potential energy incurred during the contraction
is radiated away. This is at variance with 
detailed studies which show that the radiated energy is
at best only a few percent of this \cite{haensel}. 
Most of the ``lost'' potential energy is transfered into  
internal energy (i.e. it heats the star up). 
Using results for a uniform density
sphere, one can estimate that the change in potential
energy $\delta E$ associated with a change in radius $\delta R$ is 
\begin{equation}
\delta E \approx { 3 \over 5} { GM^2 \over R^2 } \delta R
\end{equation}
Suppose that the contraction associated with a phase-transition 
in the core of a neutron star leads to $\delta R \approx 10$~m, 
and that 1\% of $\delta E$ is radiated as gravitational waves
(which does not seem too unreasonable). Then we find that
$h_{c} \approx 10^{-23}$ (assuming $f_{Hz}= 10^3$)
for a source at the distance of 
the Virgo cluster. This is probably too weak to be detected
(at least in the next few years). Still, there are several
reasons why these scenarios should not be ignored.
First of all, the event may be more violent than I have assumed
here. Secondly, a unique event from within our galaxy could 
be detectable. Given the estimated event rate (unlikely 
larger than $10^{-5}$/yr/galaxy) we 
would obviously be very lucky to see such an event, 
but the information that an observation would provide
about  physics  beyond supranuclear density
would be extremely valuable.

\subsection{The bar-mode instability}

Even though realistic neutron star
equations of state do not allow values of $\beta$ 
much larger than $0.1$ in uniformly rotating neutron stars, 
several scenarios  may lead to a compact 
star becoming dynamically unstable to the 
bar-mode. For example, since $\beta\sim 1/R$ one might expect 
a collapsing star to suffer a triaxial 
instability at some point during its 
evolution. A key parameter that determines the outcome is
the degree of differential rotation. 
In fact, the maximum attainable $\beta$
changes dramatically if the star 
is differentially rotating. This is not surprising since the 
presence of differential
rotation may lead to an increase of the mass-shedding limit
by allowing the equator to rotate slower than the
central parts of the star. 

The critical value at which the bar-mode becomes
dynamically unstable remains close to the result for 
Maclaurin spheroids, $\beta_d \approx 0.27$, for 
models with varying compressibility. 
Moreover, detailed studies suggest that the onset of 
instability is in general weakly
dependent also on the chosen differential
rotation law \cite{toman}. Having said this, there are extreme
angular momentum distributions for which $\beta_d$ becomes
very small \cite{pick}, and for which ``spiral instabilities'' 
tend to dominate. This is  interesting
as it indicates that dynamical instabilities could play 
a role also for relatively slowly rotating stars. 
This possibility is illustrated by recent 
numerical work by Centrella et al  \cite{centetal}.
Their simulations of differentially rotating $\Gamma=4/3$ 
polytropes indicate the presence of a dynamical instability
for $\beta \approx 0.14$ (i.e. similar to 
the point where secular instabilities are expected to set in). 
The simulations show that an $m=1$ mode plays a dominant
role in determining the evolution of the system. 
The nature of this instability, i.e. to what extent 
it is a generic feature, is not yet well understood.
Very recent work by Shibata et al \cite{ske02} supports the notion 
that dynamical instabilities may operate at very low values of $\beta$. 
In fact, they find that a dynamical instability may set in at
values as low as $\beta \approx 0.03$!
These intriguing results show that we still have much to learn
about the oscillations and instabilities of differentially 
rotating stars.  

In order to study the nonlinear bar-mode evolution
one must resort to large scale numerical simulations. 
Such work,  carried out over the last two decades, 
shows that the nature of the bar-mode instability 
depends on the magnitude
of $\beta$ compared to the critical value. 
For large values, $\beta >> \beta_d$, the initial exponential 
growth of the unstable mode (on the dynamical timescale) 
is followed by the formation of spiral 
arms. Gravitational torques on the spiral arms lead
to the shedding of a mass and angular momentum. 
Through this process the unstable mode saturates and the
star reaches a 
dynamically stable state. In this scenario 
gravitational waves are 
emitted in a relatively short burst \cite{houser}.
For some time it was thought that this was the generic
behaviour, but recent work \cite{new00} indicates that 
when  $\beta$ is only slightly
larger than $\beta_d$ a long-lived ellipsoidal structure 
may be formed. If this is the case, the bar-mode may
decay slowly (on the viscosity/gravitational-wave 
timescale) until the star reaches the point 
where it is seculary stable \cite{new00}. This could lead to 
a relatively long lasting gravitational-wave signal.  
Snapshots from a typical bar-mode evolution are shown in 
Figure~\ref{tsc}.

\begin{figure}[h]
\centering
\includegraphics[height=5cm,clip]{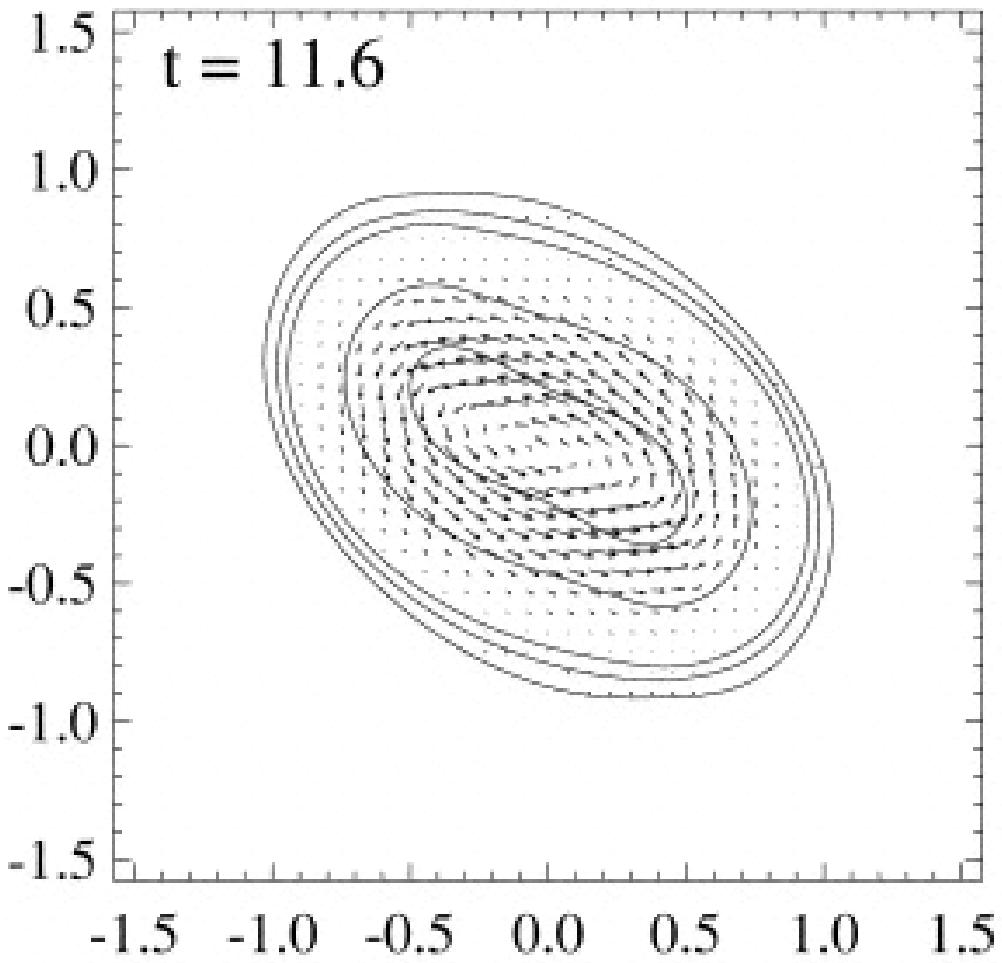}
\includegraphics[height=4.97cm,clip]{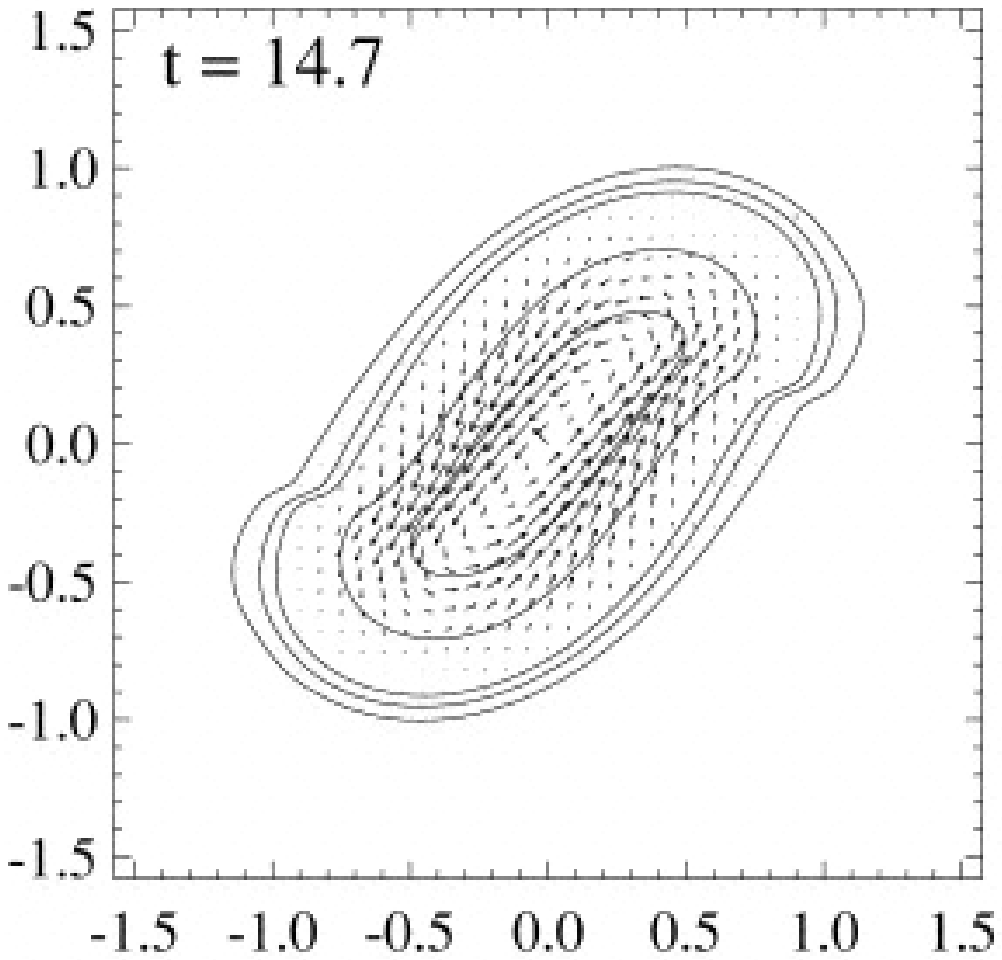}
\includegraphics[height=5cm,clip]{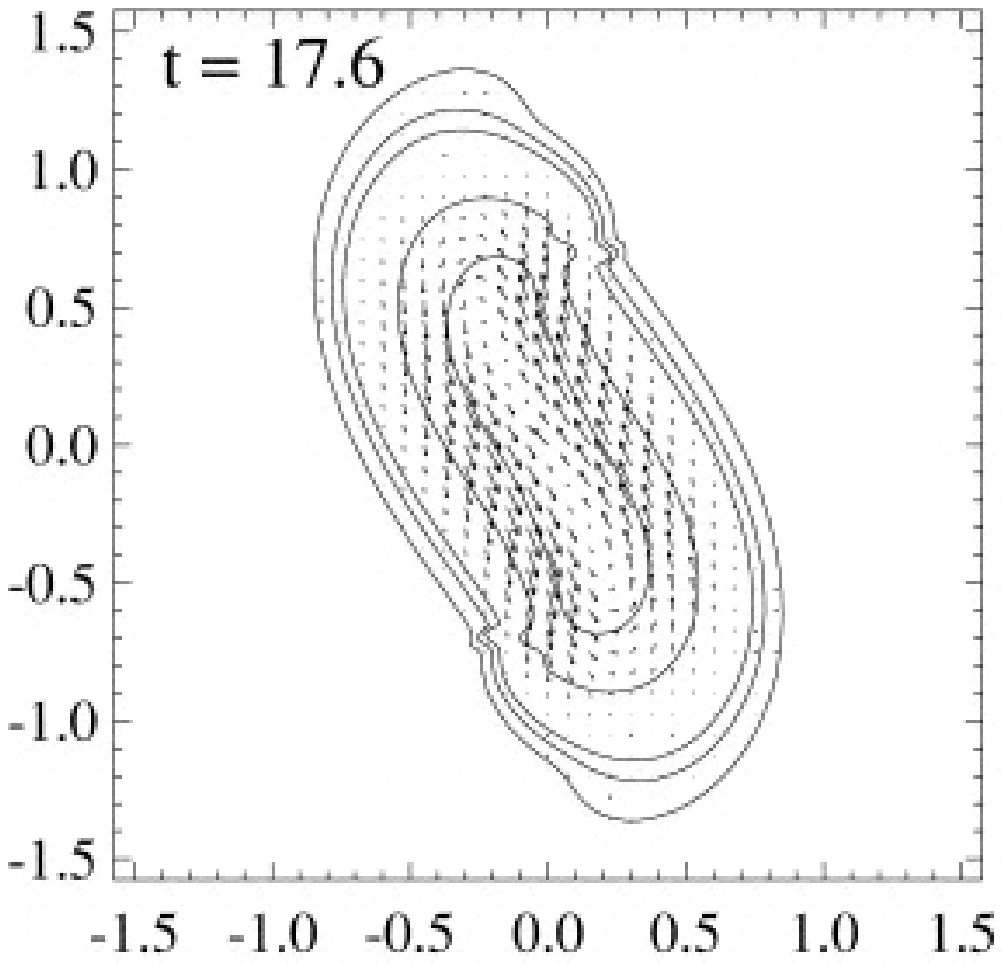}
\caption{A few snapshots from a  nonlinear evolution 
showing the development of typical bar-mode.
The initial data for the evolution is a model with uniform vortensity
and $\beta = 0.282$, i.e. slighly above $\beta_d$. 
The frames show density contours along with vectors representing the 
momenta in the equatorial plane at different times. 
The timescale indicated
in the three frames is in units of the dynamical time 
$t_{dyn} = 1/\sqrt{\pi G \rho_0}$.
[Reproduced from \cite{cazes} by kind permission from the authors.]}
\label{tsc}\end{figure}

It is straightforward to estimate the strength of the gravitational
waves emitted by a sizeable bar-mode. Let us assume that 
the mode saturates at an amplitude $\eta$ represented
by the axis ratio of the ellipsoidal structure.  
Typical values \cite{shib00} may lie in the range $\eta \approx 0.2-0.4$. 
From the standard results for a rotating solid body with a 
given ellipticity we have
\begin{equation}
{ dE \over dt} \approx -   \eta^2 {GM^2 R^4 \Omega^6\over c^5}   
\end{equation}
which leads to 
\begin{equation}
h\approx 4\times 10^{-23} \left( { \eta \over 0.2} \right) 
\left( {f \over 2 \mbox{ kHz} }\right)^2 
 M_{1.4} R_{10}^2 D_{15}^{-1}
\end{equation} 
where we have used the fact that the gravitational-wave frequency $f$ 
is twice the  rotation frequency. This estimate compares reasonably well with 
the more detailed results available in the literature 
(see, for example, Table~7 in \cite{houser}). 
A signal with this strength may be detectable
for sources in local galaxy group. 
Of course, the detectability of the signal 
is significantly improved if 
the instability leads to the formation of a persistent
bar-like structure. Should a long-lived bar form and 
last for hundreds of rotation periods, one can easily gain a 
factor of ten in the signal-to-noise ratio. Since such factors
could be crucial it is important that the
long-term evolution of the 
bar-mode instability is studied further and understood in detail. 

To perform this kind of simulations within numerical relativity (with a 
dynamical spacetime) has only recently become feasible.
Shibata et al \cite{shib00} have performed the first 
fully relativistic studies of the bar-mode problem. 
They consider models with 
varying degrees of differential rotation, and draw
conclusions that agree well with those of the Newtonian work. 
In addition, they show that general relativistic effects 
enhance the dynamical instability only very slightly
and hardly change the critical value  $\beta_d$ at all. 
Although the limited size of their numerical grid 
means that the gravitational waveforms cannot be directly
calculated, the results of  
Shibata et al \cite{shib00} seem to be in agreement
with the Newtonian estimates of the strength of
the gravitational-wave signal.

\section{Results: Secular instabilities}
\label{Sec:sec}

\subsection{The CFS instability}

As was first proved by Friedman and Schutz \cite{fs1}, the 
radiation driven instability is generic in rotating stars. 
That is, for  any given $\Omega$ one can always 
find an unstable mode (in an inviscid star). 
This is easy to see from (\ref{rotfreq}):
Regardless of the value of $\Omega$ there will always exist an
$m$ that is large enough that the associated mode  satisfies
the instability criterion, i.e. is retrograde in the rotating frame but
prograde according to an inertial observer. However, this does not 
mean that the large $m$ modes lead to the strongest instability. 
As one can readily deduce from (\ref{gwlum}) the higher order
modes tend to radiate less efficiently, and thus they will 
grow slower than the low order modes. In fact, Comins \cite{comins1,comins2}
has shown that the growth time ($t_{\rm gw}$) increases 
exponentially with $m$ for modes of the Maclaurin spheroids. 
We can  obtain a rough estimate of the growth time for
the unstable f-modes from formulas used in \cite{il2}. 
Neglecting all rotational corrections (taking 
$\alpha = \beta = \gamma=1$ in the relevant equations in \cite{il2}), 
one finds that
\begin{equation}
t_{\rm gw} (\Omega) \approx t_{\rm gw} (\Omega=0)
\left( 1 - \sqrt{{m\over 3}} { \Omega \over \Omega_K} 
\right)^{-2m-1}
\label{tfmode}\end{equation}
for the $l=m$ modes. This estimate captures the overall features
of the full numerical results and hence provides a useful illustration, 
cf. Figure~\ref{tsc}.
More detailed calculations show that 
only f-modes with $m\le 5$ are expected to grow fast enough to 
lead to an astrophysically relevant instability. 
On the other hand, the low order modes only become unstable at 
extremely high rotation rates (and the quadrupole mode may not be unstable
at all). Taking also this into consideration, one finds that 
the $l=m=4$ f-mode is the  most strongly unstable mode 
in a Newtonian star \cite{f83}. 

The situation is slightly different for the r-modes. As 
discussed previously, the r-modes are always retrograde in the 
rotating frame and prograde in the inertial frame. This means that they 
satisfy the CFS instability criterion at all rates of rotation \cite{apj,fm98}. 
In other words, the r-modes are generically unstable in rotating 
perfect fluid stars. For the 
$l=m=2$ r-mode  
one can show that the growth time is 
\begin{equation}
t_{\rm gw} \approx - 47  M_{1.4}^{-1} R_{10}^{-4} P_{-3}^6 \ \mbox{s} 
\end{equation}
for $n=1$ polytropes \cite{lom,aks99}.
It is interesting to compare this result
to the (presumably) most important
f-mode. Consider a particular $n=1$ polytropic 
stellar model with mass $1.5M_\odot$ and radius $12.533$~km
\cite{il2}, for which
the Kepler limit would correspond to a period of 0.8~ms. 
For a star spinning at this rate the r-mode would grow on a timescale
of roughly 4~s, while one finds $t_{\rm gw} \approx 
5\times 10^5$~s for the $m=4$ f-mode \cite{il2}. 
A similar comparison for other rotation rates is provided
in Figure~\ref{tsc}. These estimates indicate that
the r-mode instability is significantly stronger than 
that of the f-mode.  Having said this, one must be 
somewhat careful before drawing definite conclusions
since differential rotation could change the picture considerably.  
In particular, an unstable $m=2$ f-mode could
become competitive with the r-mode. 
For example, Lai and Shapiro \cite{lais} estimate that
the quadrupole f-mode grows on a timescale
$t_{\rm gw} \approx 1$~s for $\beta \approx 0.24$.   
One must also keep in mind that f-mode instability is strengthened 
by relativistic effects, and  that the
$m=2$ f-mode  may become unstable at reasonable rates of rotation,
cf. Section~\ref{grmodes}. An important challenge for future 
work in this area concerns the calculation of growth/damping timescales
of the f-modes of rapidly rotating fully relativistic stars. 

\begin{figure}[h]
\centering
\includegraphics[height=6cm,clip]{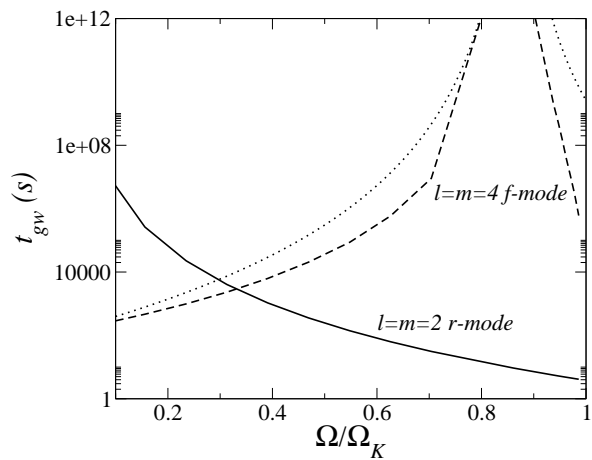}
\caption{The growth/damping timescales of the most relevant unstable modes
of a uniformly rotating $n=1$ polytrope with 
$M=1.5M_\odot$ and $R=12.533$~km. The $l=m=4$ f-mode (dashed curve)
is compared to the $l=m=2$ r-mode (solid line). We also show the 
rough estimate (\ref{tfmode}) for the f-mode (dotted curve). 
This figure captures the qualitative features of the problem (eg. that the f-mode 
becomes unstable above a critical rotation rate $\Omega_c\approx 0.85\Omega_K$), 
but comes
with several disclaimers. Most importantly, these results will be
affected by differential rotation and general relativistic effects. 
[The f-mode 
data was provided by Lee Lindblom.]}
\label{tsc}\end{figure}

\begin{figure}[h]
\centering
\includegraphics[height=6cm,clip]{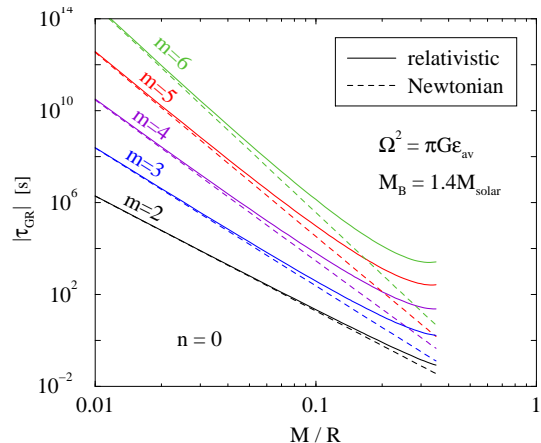}
\caption{The growth times for the inertial  modes of a 
canonical $1.4M_\odot$ relativistic barotropic (uniform density) star. 
The data is for the modes that correspond to the first few 
$l=m$  r-modes in the Newtonian limit ($M/R\to 0$) \cite{lfa02}.}
\label{grtimes}\end{figure}

To date, there have been three studies of the growth 
timescale for the unstable relativistic r-modes. Two of these, 
\cite{rk02} and \cite{yf01}, concern the non-barotropic problem
while the third was for inertial modes of barotropic stars
\cite{lfa02}. (Interestingly, the three methods used to extract the 
gravitational-wave dissipation rates are different.) 
The studies all agree that the post-Newtonian
estimates of the instability growth time are rather good. 
As can be seen in Figure~\ref{grtimes}, the growth time
$t_{\rm gw}$ approaches the post-Newtonian result as $M/R$ decreases. 
The fully relativistic timescales begin 
to deviate from the post-Newtonian ones as the star 
reaches the compactness of a typical neutron star, $M/R\approx 0.15$. 
This weakening of the instability is presumably due to the 
increased influence of the spacetime curvature as the star 
becomes more compact, and reflects
the increased backscattering of gravitational
waves.

The growth time estimates indicate that various unstable modes
may grow  fast enough to be of  significance for 
rapidly spinning neutron stars.
However, in order to assess the true relevance of these instabilities
we must also consider possible damping effects. In particular, 
an unstable mode must 
grow fast enough that it is not completely
damped out by viscosity in order to  be relevant. 
To assess the strength of  viscous damping one  typically
considers the effects of  bulk and shear viscosity. These are due to rather 
different physical mechanisms. 
At relatively low temperatures (below a few times $10^9$~K) 
the main viscous dissipation mechanism in a fluid star arises from
momentum transport due to particle scattering. In the standard
approach these scattering events are modelled in terms of a 
macroscopic shear viscosity. In a normal fluid star neutron-neutron 
scattering 
provides the most important contribution. 
The effect of the corresponding
shear viscosity is usually estimated using the viscosity 
coefficient 
\begin{equation}
\eta  = 2\times 10^{18} \rho_{15}^{9/4} T_9^{-2} {\rm g/cms} \ .
\label{shearcoeff}\end{equation}
For the r-modes, this
 leads to a dissipation time-scale 
\begin{equation}
t_{\rm sv} \approx 6.7\times10^7 M_{1.4}^{-5/4} R_{10}^{23/4} T_9^2 \mbox{ s}
\label{sv1}\end{equation}

At high temperatures (above a few times $10^9$~K) 
bulk viscosity is the dominant dissipation
mechanism. Bulk viscosity arises as
the mode oscillation
drives the fluid away from beta equilibrium.
It corresponds to an estimate of the extent to which energy is
dissipated from the fluid motion as weak interactions
try to re-establish equilibrium.
The mode energy lost through bulk viscosity is carried away 
by neutrinos. An estimate of the bulk viscosity damping rate for the 
r-modes is complicated by the fact that one must determine
the Lagrangian density perturbation \cite{aks99}.
For r-modes   this quantity vanishes at leading order
so the calculation must be carried at least to order $\Omega^2$
in the slow-rotation expansion.
In the standard case, where $\beta$-equilibrium is regulated by the 
modified URCA reactions,
the relevant bulk viscosity coefficient is
\begin{equation}
\zeta = 6\times 10^{25} \rho_{15}^2 T_9^6  
\left( {\omega_r \over1 {\rm Hz} } \right)^{-2} {\rm g/cms}\ .
\label{modurc}\end{equation}
This leads to an estimated bulk viscosity timescale for $n=1$ 
polytropes \cite{akreview} 
\begin{equation}
t_{\rm bv} \approx 2.7 \times10^{11} M_{1.4} R_{10}^{-1} P_{-3}^2 T_9^{-6} 
\mbox{ s}
\label{bulkest}\end{equation}
This result, which was obtained within the Cowling approximation 
agrees reasonably 
well with a calculation including also the perturbed gravitational 
potential \cite{lmo00}.

\begin{figure}[h]
\centering
\includegraphics[height=6cm,clip]{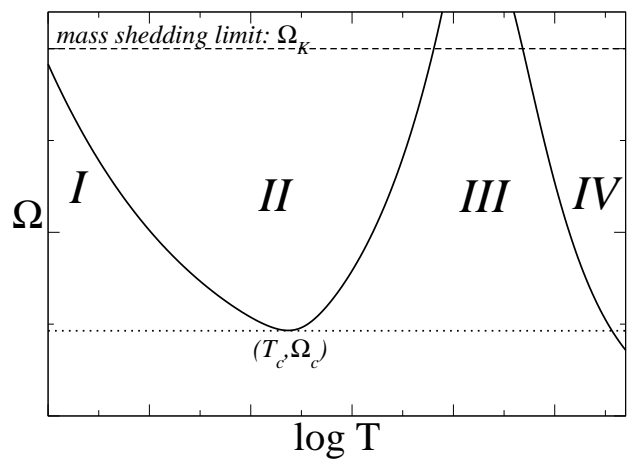}
\caption{Schematic illustration of the CFS instability window: \
At low temperatures 
(region I) dissipation due to shear viscosity counteracts the instability.
At temperatures of the order of $10^{10}$~K bulk viscosity suppresses
the instability (region III). 
At very high temperatures (region IV) the nuclear reactions that lead to 
the bulk viscosity are suppressed and an unstable mode can, in principle, grow. 
However, this region may only be relevant 
for the first few tens of seconds following the birth of a neutron star.
The main instability window is expected at temperatures near 
$T_c\approx 10^9$~K (region II).  Provided that gravitational radiation 
drives the unstable mode strongly enough the instability may 
govern the spin-evolution of a hot young neutron star.  
For the $m=4$ f-mode one finds $\Omega_c \approx 0.95\Omega_K$, while the 
$l=m=2$ r-mode leads to  $\Omega_c\approx 0.04\Omega_K$. 
The instability may change considerably
if we  add more detailed
pieces of physics, like superfluidity and the 
presence of hyperons, to the model. }
\label{window}\end{figure}

From the above estimates we can deduce that a gravitational-wave driven 
instability will only be active in a certain temperature range.
To have an  instability we need
$t_{\rm gw}$ to be  smaller in magnitude than both $t_{\rm sv}$ and 
$t_{\rm bv}$. From the  estimates above we see that 
 shear viscosity will completely suppress the 
r-mode instability at core temperatures below $10^5$~K.
This corresponds to region I in Figure~\ref{window}. 
Similarly, bulk viscosity will prevent the mode from growing 
in a star that is hotter than a few times
$10^{10}$~K. This is the case in region III of Figure~\ref{window}.
However, if the core becomes very hot (as in region IV in Figure~\ref{window}), 
then the star is no longer transparent to neutrinos and the bulk 
viscosity is strongly suppressed. This region is likely not very relevant 
in an astrophysical context as neutron stars will not remain
at such extreme temperatures for long enough that an unstable mode can grow
to a large amplitude. Finally,  in the intermediate region
II in figure~\ref{window},
 there is a temperature window  where the growth time due
to gravitational radiation is short enough to overcome the
viscous damping and drive the mode unstable. 
This general picture holds both for the unstable f-modes and the r-modes.
We find the relevant critical rotation rate, above which the
mode is unstable, by solving 
\begin{equation}
{1\over 2E} {dE \over dt} = {1\over t_{\rm gw}} + 
{1\over t_{\rm bv }} + {1\over t_{\rm sv }} = 0
\end{equation}
for a range of temperatures. 
Detailed calculations show that 
viscosity stabilizes all f-modes below $\Omega_c \approx 0.95\Omega_K$
\cite{il2}, 
while the r-modes are stable below $\Omega_c \approx 0.04\Omega_K$
\cite{lom,aks99}.
For a typical neutron star, the latter would imply that 
the r-modes are unstable at rotation periods shorter than 25~ms.
The fact that this estimate is close to the initial spin period
of the Crab pulsar inferred from observational data, $P_o\approx 19$~ms,
led to the suggestion that the r-mode instability may play a significant 
role in the spin-evolution of nascent neutron stars.
This possibility caused some excitement and spawned a multitude
of studies into the r-modes and the instability mechanism.  
Much of this work is reviewed in \cite{akreview}.
The detailed studies illustrate that this is an incredibly difficult 
problem, and that we need to understand many extremes of physics
before we can draw any reliable conclusions regarding the relevance of
the unstable r-modes. At the time of writing,  
the most important/interesting problems that need to be approached
concern: 

\underline{Differential rotation:} Neutron stars are likely to be born 
differentially rotating. This means that any serious attempt to model their
subsequent spin-evolution, eg.  driven by a gravitational-wave
instability, must be based on the oscillations of a differentially rotating
star. This is a very difficult problem, but some progress has 
nevertheless been made. Imamura et al \cite{imam95} have considered the 
onset of the secular f-mode instability for a wide range of angular momentum 
distributions. They find that differential rotation can have 
a significant effect. 
In their sample of parameter space, neutral $m=2$ f-modes occur in the 
range $0.093<\beta_s<0.14$. This means that the critical point where
the f-mode goes secularly unstable may be lowered by as much 
as 30\% by differential rotation. 
The smallest values of $\beta_s$ are found for angular momentum
distributions that are strongly peaked at the equator. 
There are as yet only one investigation into the effect that differential
rotation has on the r-modes \cite{yodif}. This study shows 
that the presence of corotation points complicates the problem 
considerably (see \cite{anna}).

\underline{Superfluid stars:}
As a neutron star cools below a few times $10^9$~K the extreme
density in the core is expected to lead to the formation of
various superfluids. The superfluid constituents play a crucial 
role in determining the dynamical properties of a rotating neutron
star. In particular, the interplay between the lattice nuclei  
and the superfluid  in the inner crust is a key agent in the
standard model for pulsar glitches. Intuitively, one would expect 
the presence of a superfluid to  have a considerable effect also on 
the various modes of pulsation.  

Studies of oscillations in superfluid stars have so far been based on 
models allowing for two distinct,  dynamically coupled, 
fluids. These represent the superfluid neutrons and the ``protons''. 
The latter can be viewed as a conglomerate of all charged components in the star.
These are assumed to be electromagnetically coupled and hence
become comoving on a very short timescale.
 Epstein \cite{E88} (and  later 
Mendell \cite{M91}) argued, using a simple counting of the fluid 
degrees of freedom, that there ought to exist
a new class of modes in a superfluid star.  
These ``superfluid'' modes have the protons moving 
oppositely to the neutrons, unlike an ``ordinary'' fluid mode that has 
the neutrons and protons moving more or less together (see \cite{acl02,prix02}
for detailed discussions). An important feature of the superfluid problem is
that a momentum induced in one of the 
constituents will cause some of the mass of the other to be carried 
along.    Because of this ``entrainment'', the flow of neutrons 
around the neutron fluid vortices (recall that a superfluid
mimics large scale rotation by forming a large number 
of vortices) will  induce a flow in a 
fraction of the protons, leading to magnetic fields being formed 
around the vortices.  But since the electrons are coupled to the 
protons on very short timescales, some electrons will track 
the entrained protons.  Mutual friction is the dissipative 
scattering of these electrons off of the magnetic fields associated 
with the vortices \cite{mend_diss}.  

Superfluid mutual friction has been shown to  suppress the 
instability of the f-mode in a rotating (Newtonian) star \cite{lm95}. 
It was originally thought to be the
dominant damping agent also on the r-modes, but a  calculation
by Lindblom and Mendell \cite{lm00} suggests that 
a typical result for r-mode
dissipation due to mutual friction is
\begin{equation}
t_{\rm mf} \approx 2\times10^5 P_{-3}^5 \mbox{ s} \ .
\end{equation}
This would mean that the r-mode instability window 
is essentially unaffected by the inclusion of mutual friction in the model.
However, $t_{\rm mf}$ is sensitive to 
changes in the entrainment parameter \cite{lm00}, and there seems to be
critical values for which the mutual friction 
timescale becomes very short. These ``resonances'' are
as yet not well understood \cite{ac01} 
and it is clear that much more detailed
studies into the  pulsation properties of superfluid neutron stars 
and the associated dissipation mechanisms are needed \footnote{This
point is further emphasized by the very recent demonstration that a so-called
two-stream instability can operate in superfluid systems \cite{twost}.}.

\underline{Ekman layers:}
In addition to the core superfluid, we need to consider
effects due to the presence of a solid crust in a mature
 neutron star. The melting temperature of the crust 
is usually estimated to be of the order of  $10^{10}$~K (for a
non-accreting  star), 
so the crust may form shortly after the neutron star is born. 
The presence of a solid crust will have a crucial effect 
on oscillations of the core fluid. If the crust is assumed to be rigid
the fluid motion must 
essentially fall to zero at the base of the crust.
One can estimate the relevance of the crust using 
viscous boundary layer theory \cite{bu00}. The region 
immediately beneath the crust
then corresponds to a so-called Ekman layer.
The thickness of the boundary layer ($\delta$) can be deduced
by balancing the Coriolis force and shear viscosity:
\begin{equation}
\delta \sim \left( {\eta \over \rho \Omega} \right)^{1/2}
\end{equation}
where $\eta$ is the shear viscosity coefficient (\ref{shearcoeff}).
After putting numbers into this relation we see that 
$\delta$ will typically be a few centimetres for
a rapidly rotating neutron star. For a core r-mode the
dissipation timescale due to the presence of the Ekman layer 
has been estimated as \cite{akreview}
\begin{equation}
t_{\rm Ek} \approx 830  T_9 P_{-3}^{1/2} \mbox{ s}  \ .
\label{tek2}\end{equation}
From this estimate one can see that a 
solid crust would have a greater influence than many other
 dissipation mechanisms. For example, one finds  
that all neutron stars with a rigid crust are stable at rotation periods
longer than roughly 5~ms, or $\Omega_c \approx 0.4$.

The crust-core interface has been the focus of 
several recent studies. These studies add further dimensions
to the problem.  The interplay between the r-modes in the 
fluid core and modes in the solid crust is particularly 
interesting. Calculations by  Yoshida and Lee \cite{crust1} 
show that  as the spin
of the star increases the  r-modes will undergo a series
of so-called avoided crossings with  modes that are due to the 
elasticity of the crust. Furthermore, Levin
and Ushomirsky \cite{crust2} have shown that the assumption of a 
rigid crust, which was made
in most estimates of the Ekman layer dissipation rate,  
is  not warranted. They show that the r-mode typically extends
into the crust, and as a consequence the true dissipation
may well be a factor of perhaps a hundred weaker than (\ref{tek2}).
Finally, Lindblom et al \cite{crust4} have discussed whether
r-mode heating at the crust-core interface may 
melt the crust. They show that this is likely 
to be the case, and argue that the outcome may lead to the 
formation of partly frozen, partly melted crust regions analogous to 
ice chunks in the Arctic. 

Despite some recent advances in our understanding of the
effects that a solid crust may have on the r-modes,  
several crucial issues remain
to be investigated in detail.
For example,  the inner crust of a neutron star (out to the neutron
drip density) will likely be permeated by superfluid neutrons. 
It is not at all clear at the moment whether one should 
expect these neutrons to be strongly pinned to the 
crust nuclei or not \cite{pbjo,cutlink}. But if the superfluid is at all
free to move relative to the crust it will 
likely lead to a weaker Ekman layer
dissipation on the r-modes.
Insights into this problem could also help shed light 
on the dynamics of pulsar glitches.

\underline{Magnetic fields:} Given that a strong magnetic field is
a key element of  neutron star physics it is somewhat surprising
that there have been few discussions of the effect that 
these fields may have on radiation instabilities. 
Particularly since the CFS instability is
not unique to gravitational radiation: Any radiative mechanism will
do, and it would not be surprising if a detailed investigation 
were to unveil interesting instabilities driven by electromagnetic radiation. 
But even if this does not turn out to be the case, 
the interplay between a large amplitude pulsation mode and  
the magnetic field is interesting. Especially since
the consequences may be observable electromagnetically. Of course, 
it is very difficult to incorporate a realistic magnetic field in a study of 
neutron star dynamics.

There have been some 
discussions of magnetic fields in connection with the gravitational-wave
instability of the r-modes. In particular concerning
the relation between an unstable mode and differential rotation. 
Spruit \cite{henk} has suggested that gravitational radiation 
reaction  induces strong differential rotation in the star. 
This leads to a winding up of the interior 
magnetic field  until  a point is reached where the field 
becomes unstable due to buoyancy. This scenario was proposed as
a model for gamma-ray bursts. 
The question whether an unstable r-mode leads to differential
rotation, and whether the mode can be prevented from growing 
by the magnetic field was  discussed by Rezzolla, Lamb and 
Shapiro \cite{rez1}. They studied the so-called Stokes drift
(due to which fluid elements undergo a secular drift when a 
wave is present in the system), the magnitude of which depends on the 
latitude of the fluid element and the r-mode amplitude. 
Key questions concern how this  
differential drift affects the magnetic field of the star, and what the
backreaction on the mode may be.  It was estimated that 
the instability could
operate in  young neutron stars ($B\sim 10^{12}$~G) and recycled ones
($B\sim10^8$~G) provided that they spin fast enough.
Just like in Spruit's model, the differential drift due to the r-mode
twists the magnetic field which
affects its strength and nature. Rezzolla et al
predict that  the r-mode instability 
generates strong azimuthal magnetic fields in young pulsars. 

Several other issues regarding the possible role of the magnetic field
were discussed by Ho and Lai \cite{hl99}. 
In particular, they attempt to
quantify the extent to which electromagnetic radiation 
from the r-modes affects the growth rate of the 
instability. They find that  the electromagnetic driving of
the mode would become competitive with  gravitational radiation for
$B\approx 10^{15}$~G.  Ho and Lai also point out that the mode-oscillations
will generate Alfv\'en waves in the magnetosphere. These would
 strengthen the instability somewhat.
Finally, Mendell \cite{menmag} has extended the crust/core
boundary layer approach to the case of a magnetized core. 
His results indicate a significant damping  of the r-modes
for magnetic fields of the order of $10^{12}$~G and stronger. 
These various results indicate 
that magnetic field effects must be included in any 
detailed r-mode model, and also point to some potentially
interesting repercussions for the instability. 

\underline{Exotic bulk viscosity:}
The presence of exotic particles in the core of a neutron star may 
lead to significantly stronger viscous damping than assumed in 
the ``standard'' instability analysis \cite{il2,lom,aks99}.
Of particular relevance may be the presence of hyperons. Jones 
has estimated that the r-mode instability
is almost completely suppressed in a star with a sizeable hyperon core 
\cite{pbj}. 
A more detailed analysis of the role of hyperons has been 
carried out by Lindblom and Owen \cite{lo02}.
Their results show that the dissipation due to hyperons is, indeed, 
overwhelming. Most importantly, the hyperon bulk viscosity 
has the same temperature dependence as the shear viscosity, and
thus it limits an instability at low temperatures.
The available estimates suggest that the instability 
is unlikely to be significant for neutron stars with significant hyperon
fractions. However, it is important to emphasize the many 
uncertainties concerning exotic neutron star cores. For example, 
hyperons would act as a very efficient refrigerant. In fact, 
a neutron star with a hyperon core should  cool rapidly
to temperatures much lower than those suggested
by observational data. This discrepancy can be avoided if the 
hyperons are superfluid and the relevant nuclear reactions are
suppressed. But if this is the case then the bulk viscosity is 
also suppressed, and in addition one must discuss the 
role of the additional degree(s) of freedom that follows
with having a star with superfluid components.  

Interestingly, a strong viscous damping could work in favour of
the r-modes as a gravitational-wave source. This is illustrated
by the case of strange stars. The observational evidence for the
existence of strange stars is tenuous, but they may exist if
strange matter is indeed the most stable form of matter
at high density. 
As was first pointed out by Madsen \cite{jm99},
the r-mode instability is affected by the fact that
the bulk viscosity of strange matter 
is many orders of magnitude stronger
than its neutron star counterpart.
This means that the main instability window is located at 
comparatively low temperatures in a strange star. 
In Figure~\ref{window} we would have 
$T_c\approx 10^7$~K and $\Omega_c \approx 0.25\Omega_K$. 
The fact that the instability window is shifted to 
lower temperatures means that an accreting strange star may 
become a persistent gravitational-wave source once it 
reaches the spin-rate where the r-modes become unstable.
This possibility will be discussed further in the next section.

\underline{Nonlinear saturation mechanisms:}
To model the gravitational waves associated with a secular instability
we need to understand how an unstable mode evolves and
what the ``backreaction'' on the bulk of the star is. 
This is a very difficult problem. The early growth phase 
of an unstable mode can obviously be described by linear theory, 
but we need to model what happens when the 
mode enters the nonlinear regime. 
Intuitively, one would expect the growth of the mode to be halted
at some finite amplitude. It seems
plausible that the excess angular momentum will be radiated away 
as the mode saturates, and that the star will spin down as a consequence.
This general picture is supported by 
studies of instabilities in rotating 
ellipsoids \cite{miller,detweiler1,detweiler2}.
In particular, 
Lai and Shapiro have discussed gravitational-wave
signals associated with unstable ellipsoids in great detail. 
Their evolutions are driven by an unstable f-mode, and they 
argue that the characteristic gravitational-wave 
amplitude is
\begin{equation}
h_c \approx 1.2\times10^{-22} M_{1.4}^{3/4} R_{10}^{1/4} f_{Hz}^{1/2} 
D_{15}^{-1} 
\label{fest}\end{equation}
As the star spins down, the frequency of the signal varies 
in such a way that $f_{Hz} \approx 10^3 \rightarrow 0$, 
i.e. it moves through the region where  
groundbased interferometers will be the most sensitive.
However, equation (\ref{fest}) is likely overly optimistic because it assumes that 
the spin-down is governed by an $m=2$ mode. As discussed above, 
this mode leads to the fastest instability growth time, but  it is not 
 clear that the $m=2$ f-mode will be unstable in a 
realistic (compressible) model. Having said that, one should not rule out the 
possibility since both General Relativity and differential 
rotation tend to destabilize the f-mode. 

As far as the unstable r-modes are concerned, most models
to date have been phenomenological. The first such model, developed 
by Owen et al \cite{owen},  was based on expressing the conservation of energy and 
angular momentum as a system of evolution equations for the bulk rotation 
rate $\Omega$ and the mode-amplitude $\alpha$. The latter was defined by 
\begin{equation}
\delta \vec{v} \approx \alpha \Omega R 
\left( { r \over R} \right)^l \vec{Y}_{ll}^B e^{i\omega_r t}
\label{normalise}\end{equation}
In order to account for nonlinear mode-saturation it was simply assumed that 
$\alpha$ stopped growing at a suitably large value.
Qualitatively, such evolutions would be in accordance with the 
results for
the analogous problem for ellipsoids \cite{miller,detweiler1,detweiler2}.

A key question concerns what the maximum value of $\alpha$ is. 
In the first studies, it was  assumed that 
$\alpha$ could reach values of order unity. This would lead to a 
newly born neutron star spinning down significantly on a timescale
of a few weeks-months. The resultant gravitational-wave signal 
would be detectable by LIGOII for sources in the Virgo cluster \cite{owen}.
The mode-saturation has recently been 
investigated in two different ways. 

The first studies of this problem were based on
fully nonlinear hydrodynamics.  
Stergioulas and Font \cite{sfevol} carried out a numerical
evolution of the exact relativistic Euler equations on a background
spacetime using a  large-amplitude r-mode as  initial data.  
They found that there was no apparent energy transfer into
other modes until $\alpha$ was substantially
larger than unity.  This result was confirmed
by evolutions of the full Newtonian hydro equations
 by Lindblom, Tohline and Vallisneri \cite{ltevol,ltev2}.
 Because the actual growth time due to radiation reaction is
impractically long, they studied the maximum amplitude for a
perturbation driven by an enhanced radiation reaction force \cite{rez2}. 
They again found saturation  at an amplitude large compared to unity.
The limit appeared to be set by a shock wave, a dramatic 
wave-breaking on the
star's surface.
A clear implication of these numerical evolutions was that coupling to
other low-order modes did not set a stringent limit on the r-mode amplitude.
At least not within the timescale of the evolutions (about ten 
rotational periods). To what extent these simulations represent 
the true physical behaviour was not clear, given the 
limitations in resolution and evolution time.
Ideally, one would want to study the onset of instability 
and follow the mode through to saturation and see what effect
the instability has on the spin of the star. Computationally, 
this means that one would need a numerical evolution that resolves
the mode-oscillation (on a millisecond timescale)
and tracks the star through hundreds of seconds. 
This is not possible given current technology. 

A complementary way to address questions concerning nonlinear mode-coupling
proceeds via second-order perturbation
theory.  By omitting higher-order couplings one can reduce the nonlinear
evolution to a set of coupled ordinary differential 
equations for the amplitudes
of the various modes of the system.  The development of this 
perturbation theory for rotating stars was a  major undertaking,
which was recently completed by Arras et al \cite{arras}.
They find that a strong resonant coupling to short wavelength 
inertial modes would lead to r-mode saturation at an 
amplitude
\begin{equation}
\alpha \approx 8\times 10^{-3} P_{-3}^{-5/2} 
\end{equation}
for a canonical star.
This value is two orders of magnitude smaller (for rotation rates near 
the Kepler limit) than that used in the early studies of 
the r-modes \cite{owen}. This 
means that the effect of the instability would be much less dramatic.
 
At first sight it may seem as if the outcome of the perturbation 
calculation contradicts the 
nonlinear evolution results. However, the work of Arras et al \cite{arras}
is consistent with the nonlinear evolutions
in the sense that the coupling they find to low-order modes would lead to a 
saturation amplitude of order unity. It is the coupling to
many short-wavelength modes, with frequencies comparable to that of the
r-mode, that saturates the mode at a small amplitude. 
Given the coarse resolution of the numerical grids one would not
expect these short wavelength oscillations to show up 
in the nonlinear evolutions. This interpretation is supported by 
the most recent fully nonlinear numerical evolution,
due to Gressman et al. \cite{gress},
which shows  that increasing the resolution yields
rapid nonlinear decay of the mode at decreased amplitude.
This work points the way to future studies of the problem. The challenge
is to test the saturation amplitude predicted by perturbation theory, 
perhaps via high resolution numerical evolutions.
At the present time, the 
r-mode saturation remains an issue that 
requires further detailed consideration.

\subsection{The viscosity driven instability}

As discussed in Section~\ref{el_sec}, viscosity can drive 
an instability in prograde moving oscillation modes of a
spinning star \cite{roberts}. This instability sets in at the point
where the rotational corrections to the mode frequency 
are such that the mode has zero frequency in the rotating frame, 
cf. (\ref{ecrot}). A viscosity unstable mode 
lowers the kinetic energy of the
star by converting it into heat. In contrast to the radiation driven 
instability, this instability is not generic. 
The reason for this is that the quadrupole f-mode is the first
to go unstable. Higher multipoles require faster 
rotation to reach their respective instability points. 

By an interesting ``coincidence'' the critical 
rotation rates for the viscous
instability and the radiation driven CFS 
instability are identical in a rotating Newtonian ellipsoid.
Both instabilities become active at $\beta_s \approx 0.14$. 
This result no longer holds for compressible stars.
In fact, one finds that the viscous instability will not
be present unless the equation of state is unexpectedly stiff. 
For example, in Newtonian polytropes it 
is present only if $n<0.808$ \cite{tassoul}. 
The viscosity driven instability is also not
favoured by General Relativity. Relativistic effects tend
to stabilize the modes that are susceptible to the viscosity 
instability, and one can show that the instability 
may operate in $1.4M_\odot$ relativistic stars 
only if $n\le 0.67$ \cite{skinner}.  
The neutron star equation of state is
expected to be significantly softer than this. Nevertheless, the 
first relativistic study of the problem \cite{bonna} indicates that
a subset of the proposed equations of state allow the 
instability to operate near the mass-shedding limit.  
One should also not rule out the 
possibility that rapidly spinning strange stars may undergo 
the viscous instability \cite{gour}. 

There have only been
two studies of the possible gravitational-wave signal 
from a star undergoing the viscous instability. 
Almost twenty years ago, Ipser and Managan \cite{im84} 
estimated that an energy of $10^{-4}M_\odot c^2$ would be radiated
through a signal with $f_{Hz} \approx 1250$.
The signal would have very narrow bandwidth (a few Hz) since the 
star would not migrate far from the bifurcation point
(cf. figure~\ref{ellipse}). This is a reasonable assumption
since the viscosity driven instability is likely to exist 
(if at all) only near the mass-shedding limit. 
These conclusions were confirmed, and somewhat refined, by
Lai and Shapiro \cite{lais} who considered unstable ellipsoids.

However, it seems likely that damping due to gravitational-wave 
emission will suppress the viscous instability in all astrophysical 
neutron stars. The argument for this is quite simple. 
Just like in the case of the CFS instability the mechanism 
that drives the unstable mode must overcome all damping agents. 
In this case the viscous timescale associated with the unstable 
mode must be shorter
than the gravitational-wave damping timescale. For the modes that 
are the most likely to suffer the viscosity driven instability
(the prograde moving quadrupole f-modes) the various
timescales will be weakly affected by the stellar rotation. 
This means that we can use the non-rotating values to estimate 
when the viscous timescale is shorter than that of gravitational-wave
damping. For the shear viscosity we have \cite{cutlin}
\begin{equation}
t_{\rm sv} \approx { 1\over 5} { \rho R^2 \over \eta}
\end{equation}
while the gravitational-wave damping should be adequately 
approximated by (\ref{tgwnon}).
Combining these estimates we find that we must have 
$T<2\times10^4$~K in order to have $t_{\rm sv}<t_{\rm gw}$. 
This shows that the viscosity driven instability may only operate in 
extremely cold neutron stars. However, neutron stars as
cold as this are unlikely to exist in the Universe as accretion 
from the interstellar medium may prevent cooling
significantly below $10^5$~K \cite{owpri}.

\section{Astrophysical scenarios}

It is not difficult to argue that neutron stars 
should be born rapidly spinning. By simply assuming 
conservation of angular momentum during the collapse of 
a solar-mass core from a radius of more than 1000~km to 
perhaps ten km, one finds that neutron stars ought to be born 
rotating as fast as they possibly can, i.e. near the break-up limit.
However, this may be oversimplifying the problem greatly. 
In fact, Spruit and Phinney \cite{sp98} have argued
that magnetic locking between the core and the envelope of
the progenitor star may prevent the 
collapsing core from spinning rapidly.
In their model the main stellar rotation is due to the  
``kick'' mechanism that may also cause a large linear
momentum. Consequently, one would expect only a small subset of neutron stars
to be born with spin periods shorter than (say) 10~ms.  

Observational data regarding the initial spin rate of 
young pulsars is not reliable enough  to 
piece together a coherent picture \cite{fallback}. 
The best studied case is the Crab pulsar, 
whose initial period is estimated (assuming the standard
magnetic braking model) to have been about 19~ms. 
The recently discovered young 16 ms X-ray pulsar in the supernova 
remnant N157B should have been born spinning  faster, but it still 
probably had an initial period no 
shorter than a few ms \cite{marshall}.
These estimates should be compared to 
the shortest known period 
of a recycled pulsar of 1.56~ms, and with the 
theoretical lower limit on the period of about 0.5 to 2 ms \cite{nslr}, 
depending on the equation of state. 

One possible explanation for the absence
of young pulsars spinning near the Kepler limit
could be that rotational instabilities
play a significant role in the spin-evolution of nascent neutron stars,
leading to a loss of angular momentum during or immediately 
after the initial collapse. In this respect the first estimates for the 
r-mode instability looked promising \cite{lom,aks99}. Phenomenological
spin-evolution models suggested that
the unstable r-modes 
would be able to spin a newly born neutron star down to a rotation rate 
of roughly 20~ms, in good agreement with the observations of 
the Crab pulsar. It was also pointed out \cite{aks99} that
the r-mode instability might have the consequence that  young neutron stars
can only reach  
rotation periods shorter than (say) 3-5~ms if they are
recycled by accretion in 
a binary system.  The 
alternative model, that these stars are formed by accretion-induced 
collapse of a white dwarf, was not  consistent
with the r-mode results.  The collapse would simply form
a star hot enough that it would be expected to spin down 
because of the instability. Given more detailed
studies of the relevant dissipation and mode-saturation mechanisms
\cite{akreview}
it is not clear to what extent the original r-mode scenario
is still viable. The emerging picture is complex, and the outcome 
 depends crucially on physics that is not well understood.

There have not been many fully nonlinear studies aimed at establishing 
whether rotational instabilities play a role
in gravitational collapse scenarios. The reason for this 
is that  three dimensional studies of rotating 
core collapse are still prohibitively difficult (given current computational
technology). 
The issue was, however, discussed by Rampp et al \cite{rampp} who
obtained evidence that instabilities are unlikely to be 
 relevant during the collapse event itself. 
They found that there were events in the Zwerger-M\"uller
collapse catalogue \cite{zwmull} that led to $\beta>\beta_d$, 
and therefore in principle the presence of a dynamically 
unstable bar-mode, 
at the time of core bounce. But since the core tended to 
reexpand before the unstable mode had time to grow to a 
large amplitude no enhancement in the gravitational-wave
signal was observed. This may well be a generic feature, 
and we should perhaps not expect instabilities to 
emerge from collapse events involving a strong bounce (at least not
until the remnant settles down).
The strongest gravitational-wave signal estimated from the 
Zwerger-M\"uller data would correspond to a radiated energy of
$E\approx 8\times10^{-8}M_\odot c^2$ with a characteristic
frequency of a few hundred Hz. Such a signal is unlikely to be detected
from sources beyond the local group of galaxies. 
The event rate for such sources is unfortunately 
very low, with only a few supernovae expected in 100 years
in this volume of space.

Other studies have provided more 
promising (from the gravitational-wave point of view)
results. Houser, Centrella and Smith \cite{hcent}
studied the development of unstable bars which grow while the 
collapse is stalled by the centrifugal force. Their results
suggest that as much as $E\approx 10^{-3}M_\odot c^2$ could be 
radiated as gravitational waves. 
However, since the typical frequency of the waves is in the 
kHz range,  this is probably not enough 
to make the signal 
detectable from events in the Virgo cluster. 
Furthermore,  the study comes with a 
significant disclaimer since the entire collapse is not followed. 
The initial configuration chosen for the simulation may 
be one that is never actually reached in a ``realistic'' core collapse.
This is an issue that needs to be investigated in more detail 
 in the future: We need to understand what the ``physical''
region of the parameter space is.

It has been suggested that gravitational collapse
may under some circumstances be halted by the centrifugal force 
and/or thermal pressure
 between white dwarf
and neutron star densities. In this region there are no stable
equilibrium configurations. In the original scenario, 
the formed object was assumed to be dynamically
but not secularly stable. This would lead to a phase 
during which angular momentum is lost 
due to gravitational-wave driven instabilities.
The system slowly spins down and contracts as 
angular momentum is lost  until a neutron star is formed.  
Since this corresponds to a failed supernova
this non-explosive avenue for neutron star formation 
became known as a ``fizzler''. However, Shapiro and Lightman
\cite{sl76} showed that the scenario is unlikely to work for a 
star governed by a standard ``cold'' equation of state.
The fizzler state will only exist for a 
part of the evolution --- the system will become 
dynamically unstable before it reaches neutron star densities. 
Recent work by Hayashi et al \cite{haya1} revives the idea by 
showing that the fizzler scenario could work for hot stars.
A rotating configuration that is described by a hot equation 
of state can be dynamically stable because of the thermal
pressure associated with a large lepton fraction. 
A preliminary study \cite{haya2} indicates that  
relativistic effects tend to make the formation of 
fizzlers more difficult, but they cannot be ruled out.

In an interesting recent paper, Imamura and Durisen \cite{imdu}
have  argued that the fizzler evolution it likely
to be dominated by dynamical, rather than secular, instabilities. 
This result is based on the notion that 
deleptonization and cooling leads to constraction of the 
spinning configuration. As a consequence, the 
system approaches the dynamical instability point. 
Imamura and Durisen estimate the energy radiated 
as gravitational waves during the fizzler phase to be
\begin{equation}
{ \Delta E \over M c^2} \sim 10^{-3} 
\left({M \over M_\odot} \right) \left( 
{P \over 10\mbox{ ms}}\right)^{-1}
\end{equation}
where $P$ is the rotation period of the marginally stable state. 

Fryer et al \cite{fryer} have recently argued that the formation 
of medium size black holes from first generation stars could 
lead to fizzler-type graviational-wave signals. The key idea is
that angular momentum delays collapse of a $\sim 50 M_\odot$ 
core at a radius $R\sim 1000$~km. If a secular instability 
develops (as in the original fizzler scenario) 
then it could lead to a significant amount of
energy
being released as gravitational waves. 
This may lead to a
signal strain of $h\sim 10^{-21}/d$~Gpc, where the typical distance
to the source will be greater than  5~Gpc. The frequency
of the waves will be $f_{Hz}\sim 10^{-2}$ and thus the signal 
could be of relevance for LISA.
A similar scenario was proposed by New and Shapiro \cite{new01}. They 
consider the collapse of supermassive stars, with $M\approx 10^6M_\odot$. 
Should the collapse of such stars be stalled at a 
radius of $10^{17}$~m, a secular instability could lead to a 
gravitational-wave signal of $h\approx 10^{-15}$ for a source at 
1~Gpc. The corresponding wave-frequency would again be $f_{Hz}\approx 10^{-2}$. 

All the fizzler-type scenarios are interesting, and could
well lead to relevant gravitational-wave
signals. But it is clear that much more thought needs to 
go into the relevant stability criteria and 
the various viscous dissipation mechanisms that may 
affect the evolution of the proposed secular/dynamical instabilities. 
We also need to establish whether rotating core collapse can lead to 
the formation of objects in the range $\beta_s < \beta < \beta_d$.  
So far we may only have scratched
the surface of this challenging problem.


A secular gravitational-wave driven instability may also play 
a role in accreting systems. 
The possibility that gravitational waves from 
CFS unstable f-modes could balance the accretion torque, and hence
halt the associated spin-up, was first discussed by
 Papaloizou and Pringle \cite{papaloizou} and  
Wagoner \cite{wagoner}. 
The analogous scenario for the r-modes was 
analysed in detail by Andersson, Kokkotas and 
Stergioulas \cite{akst99} (see also  Bildsten \cite{bildsten98}).
It was originally thought that an accreting star in which 
an unstable mode could be excited to a significant level
would reach a spin-equilibrium where gravitational-wave emission balances the 
torque. 
Should this happen, the neutron stars in Low-Mass X-ray Binaries (LMXBs) 
would be promising
sources for detectable gravitational waves. 
For example, if gravitational-wave emission
provides a limit on the spin of Sco~X1, which is the  strongest 
X-ray source in the sky, and  the 
average accretion rate onto the neutron star is $\dot{M}\approx 3\times
10^{-9}M_\odot/\mbox{yr}$ then we can deduce that
$h\approx 3.5\times10^{-26} \rightarrow h_c \sim 10^{-21}$
(for $P\approx 4$~ms and $f_{Hz} \approx 330$)
after two weeks worth of signal has been accumulated \cite{akst99}.
In other words, this kind of source ought the be 
detectable by large-scale interferometers (perhaps requiring 
a narrowbanded advanced detector).

However, the original scenario may not work for neutron stars
\cite{henk,levin}.
In addition to generating gravitational waves that dissipate 
angular momentum from the system, the unstable mode
will heat the star up (via the shear viscosity).
If the viscosity gets weaker as the 
temperature increases (as is likely to be the case for an accreting  
neutron star
since the core temperature is expected to be $10^8$~K, which is smaller than 
$T_c\approx 10^9$~K, cf. Figure~\ref{window}), the mode-heating triggers a
thermal runaway and in a few weeks/months the star may
spin down to a 
comparatively low rotation rate. This could 
 rule out the r-modes in galactic LMXBs as a source
of detectable gravitational waves, since they will 
only radiate for a tiny fraction of the systems
lifetime (although see \cite{Heyl}). 
Interestingly, one can show that the instability
operates in a different way in strange
quark stars \cite{jm99}. Because of the significantly
stronger bulk viscosity the r-mode instability 
window shifts towards lower temperatures. This means that, once the onset of 
instability is reached,  an accreting strange star
will reach a quasi-equilibrium state on a timescale of about a year.
Hence, the r-modes in an accreting  strange star may
emit a persistent gravitational-wave signal
that would provide unique evidence for the existence
of such stars \cite{spaper}. This is an exciting possibility since a 
proof of the existence of strange stars in the Universe would
put useful constraints on the parameters of QCD.
Very recent results \cite{wag02} seem to indicate that 
accreting neutron stars with a superfluid hyperon core may 
evolve in a similar way, further strengthening the 
case for the r-mode instability to be relevant in LMXBs. 

\section{Concluding remarks}

\begin{figure}[h]
\centering
\includegraphics[height=8cm,clip]{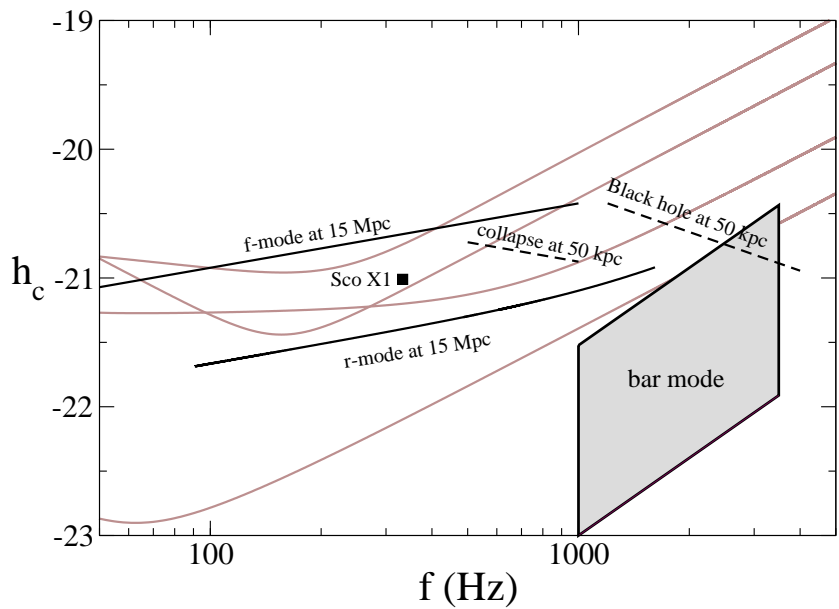}
\caption{This figure provides a (quite optimistic) 
summary of the gravitational-wave
sources discussed in the article. The effective (dimensionless) gravitational-wave strain
$h_c$ is compared to the expected sensitivity of the new generation of interferometers
(from top to bottom at the right end of the diagram: GEO600, LIGOI, VIRGO, LIGOII).
The various source estimates correspond to: r-mode spindown of a nascent neutron star
(assuming the mode can grow to a large enough amplitude and that it survives saturation
etcetera), 
the analogous f-mode spindown (assuming that the $m=2$ mode is unstable), 
gravitational collapse
radiating only an energy equivalent to $8\times10^{-8}M_\odot c^2$, and black hole 
formation (quadrupole QNM ringing assuming an energy of $10^{-3}M_\odot c^2$).
The bar-mode instability is represented by the grey region, where the upper limit corresponds 
to a source at 50~kpc and the lower limit to 15~Mpc (the Virgo cluster). 
Finally, the square indicates an accreting star which is prevented from 
spinning up by an unstable r-mode (assuming parameters relevant for Sco X1).   }
\label{hdiag}\end{figure}

The aim of writing this article was to provide an overview
of what we currently know about dynamical and secular 
instabilities in relativistic stars as gravitational-wave sources. 
There are many exciting possibilities (the various estimates are 
summarized in Figure~\ref{hdiag}) 
and it would be unwise to rule out \underline{any} of the proposed
scenarios given our current (lack of) knowledge of the detailed physics.   
For example, in the case of the r-modes
the most recent evidence suggests that i) the gravitational-wave
driven instability may be completely suppressed in a 
star with a sizeable hyperon core, and  ii) 
mode-coupling to very short wavelength inertial modes could 
lead to saturation at  low amplitudes. 
To what extent these results represent the ``final say'' on these
issues is  difficult
to assess.  First of all, the reliability of the relativistic mean-field
equations of state used in the hyperon model is 
doubted by some nuclear physicists \cite{akmal}, and (if we want to take an 
extreme point of view) it 
is not clear that hyperons will be present 
at all in a neutron star. Secondly, the mode-coupling estimates
were based on a simple barotropic neutron star model.
At some level this assumption will be 
an oversimplification.  For example, what happens to the mode-coupling 
if the neutron star core is superfluid?
It is also important to realize that, even though the main focus 
of attention has been on
the r-modes in the last few years, the secular instability of the f-modes 
is by no means dead and buried. The current Newtonian evidence may 
suggest that this instability is less 
relevant for uniformly rotating stars, but we know that 
this will change when we account for differential rotation and
relativistic effects. Most importantly, if the $m=2$ f-mode becomes unstable
it may grow on a timescale comparable to that of the r-mode. 
There are also uncertainties concerning the dynamical bar-mode 
instability.  Key questions concern whether configurations with $\beta>0.27$ 
will ever occur in a realistic collapse scenarios, and
whether the recently discovered instabilities that operate at much lower 
values of $\beta$ are of  astrophysical significance.

If we want to have 
an accurate decription of the 
emerging gravitational-wave signals  from these scenarios we need to
improve our theoretical models considerably.
These are wonderful  problems that require
an understanding of many extremes of physics
for their solution. The question is if it is realistic
 to expect us to be able to actually ``solve them''.
In many ways it seems likely  that we will need
observational data to put constraints on our
theoretical models. As the
new interferometers  may open the gravitational-wave
window to the Universe in the next few months, 
it is appropriate to ask to what extent 
the various theoretical models are falsifiable by observations.
For example, given the current ideas about the r-mode
instability one might predict that the mechanism
may not operate in an accreting neutron star, but that 
it could lead to an accreting strange star emitting 
a more or less persistent gravitational-wave signal. 
The detection of such a signal from a galactic LMXB might 
therefore indicate the presence of a  star governed
by an ``exotic'' equation of state. 
Similarly, a newly born neutron star may not undergo 
a significant r-mode instability phase if it is massive enough
to have a large hyperon core. But if hyperons are
not present, or the relevant nuclear reactions are
suppressed by the hyperons being superfluid, then 
the instability may play a role. There are many such
comparisons that one can make, and it is clear that 
they could provide highly relevant insights into
physics at extreme densities. 
Such results would be very exciting and would 
truly herald in the new era of ``gravitational-wave astronomy''.

\section*{Acknowledgements}

Generous support from a  Philip Leverhulme Prize fellowship is gratefully 
acknowledged. The authors work was supported by 
 PPARC  through grant number PPA/G/S/1998/00606, and the  EU programme 
``Improving the Human Research Potential and the Socio-Economic
Knowledge Base'' (research training network contract HPRN-CT-2000-00137).
It is a pleasure to thank the members of this EU research network
for stimulating discussions of a variety of topics.
The hospitality of the Center for Gravitational-Wave Physics at 
Penn State University, where the final touches on this review were made, 
was also much appreciated.

\vspace*{1.5cm}

\def\prl#1#2#3{{ Phys. Rev. Lett.\ }, {\bf #1}, #2 (#3)}
\def\prd#1#2#3{{ Phys. Rev. D}, {\bf #1}, #2 (#3)}
\def\rmp#1#2#3{{ Rev. Mod. Phys.}, {\bf #1}, #2 (#3)}
\def\plb#1#2#3{{ Phys. Lett. B}, {\bf #1}, #2 (#3)}
\def\prep#1#2#3{{ Phys. Reports}, {\bf #1}, #2 (#3)}
\def\phys#1#2#3{{ Physica}, {\bf #1}, #2 (#3)}
\def\jcp#1#2#3{{ J. Comput. Phys.}, {\bf #1}, #2 (#3)}
\def\jmp#1#2#3{{ J. Math. Phys.}, {\bf #1}, #2 (#3)}
\def\cpr#1#2#3{{ Computer Phys. Rept.}, {\bf #1}, #2 (#3)}
\def\cqg#1#2#3{{ Class. Quantum Grav.}, {\bf #1}, #2 (#3)}
\def\cma#1#2#3{{ Computers Math. Applic.}, {\bf #1}, #2 (#3)}
\def\mc#1#2#3{{ Math. Compt.}, {\bf #1}, #2 (#3)}
\def\apj#1#2#3{{ Astrophys. J.}, {\bf #1}, #2 (#3)}
\def\apjl#1#2#3{{ Astrophys. J. Lett.}, {\bf #1}, #2 (#3)}
\def\apjs#1#2#3{{ Astrophys. J. Suppl.}, {\bf #1}, #2 (#3)}
\def\acta#1#2#3{{ Acta Astronomica}, {\bf #1}, #2 (#3)}
\def\sa#1#2#3{{ Sov. Astro.}, {\bf #1}, #2 (#3)}
\def\sia#1#2#3{{ SIAM J. Sci. Statist. Comput.}, {\bf #1}, #2 (#3)}
\def\aa#1#2#3{{ Astron. Astrophys.}, {\bf #1}, #2 (#3)}
\def\apss#1#2#3{{Astrop. Sp. Sci.}, {\bf #1}, #2 (#3)}
\def\mnras#1#2#3{{ Mon. Not. R. Astr. Soc.}, {\bf #1}, #2 (#3)}
\def\prsla#1#2#3{{ Proc. R. Soc. London, Ser. A}, {\bf #1}, #2 (#3)}
\def\ijmpc#1#2#3{{ I.J.M.P.} C {\bf #1}, #2 (#3)}

\end{document}